\documentclass[11pt]{article} 

\usepackage{amsfonts,amssymb,amsmath,latexsym,xfrac,ae,aecompl}

\newcommand{\FFF}{\vspace*{\bigskipamount}}

\newcommand{\BBB}{\vspace*{-\bigskipamount}}

\newcommand{\cO}{\mathcal{O}}

\newcommand{\mE}{\mathbb{E}}

\newcommand{\Paragraph}[1]{\BBB\paragraph{#1}}
\newcommand{\remove}[1]{}

\newlength{\pagewidth}
\newlength{\captionwidth}
\newcommand{\qed}{\hfill $\square$ \smallbreak}

\newenvironment{myquote}
{\begin{quote}\normalsize}{\end{quote}}

\begin{document}

\baselineskip           	3ex
\parskip                	1ex

\title{Randomized Communication in Radio Networks~\footnote{This work was published as~\cite{Chlebus-chapter-2001}. It was partly prepared while visiting the University of Liverpool.
}\FFF\FFF\FFF}

\author{Bogdan S. Chlebus~\footnote{Department of Computer Science and Engineering, 
University of Colorado Denver, 
Campus Box 109, 
PO Box 173364, 
Denver, CO 80217.}}

\date{}

\maketitle

\vfill

\begin{abstract}
A communication network is called a radio network if its nodes exchange
messages in the following restricted way.
First, a send operation performed by a node delivers copies of the same 
message to all directly reachable nodes.
Secondly, a node can successfully receive an incoming message only if exactly
one of its neighbors sent a message in that step.
It is this semantics of how ports at nodes send and receive messages
that defines the networks rather than the fact that only radio waves 
are used as a medium of communication; but if that is the case then 
just a single frequency is used.
We discuss algorithmic aspects of exchanging information in such networks,
concentrating on distributed randomized protocols.
Specific problems and solutions depend a lot on the topology of the 
underlying reachability graph and how much the nodes know about it.
In single-hop networks each pair of nodes can communicate directly.
This kind of networks is also known as the multiple access channel. 
Popular broadcasting protocols used on such channels are Aloha and the
exponential backoff.
Multi-hop networks may have arbitrary topology and packets need to be 
routed hopping through a sequence of adjacent nodes.
Distributed protocols run by such networks are usually robust enough not to 
expect the nodes to know their neighbors.
These ad-hoc networks and protocols model the situation when nodes 
are mobile and do not rely on a fixed infrastructure.
\end{abstract}

\vfill

~

\thispagestyle{empty}
\setcounter{page}{0}

\newpage

\section{Introduction}

The advent of new technologies in both computers and telecommunication
has caused a proliferation of computer networks.
The ubiquitous cellular phones and portable computers have made wireless 
communication and mobile computation part of our daily experience.

Computer/communication networks are collections of information-processing 
nodes which communicate among themselves.
Nodes are often independent units and the purpose of
communication is to carry out distributed computation by sharing 
distributed resources.
Such networks need communication protocols that are versatile enough 
to handle arbitrary patterns of communication and bursty traffic.

Various taxonomies of communication networks are possible.
One of them is based on scale, that is, the size and distance among nodes;
most popular categories are {\em local area networks\/} (LANs),
like all the computers on a campus, and {\em wide area networks\/} (WANs),
like all the hosts of academic institutions in a country.
Another classification may be based on the technology used for transmission;
nodes may be connected by copper wire or optical fiber or a network may be
wireless.
Yet another distinction is by the criterion if all the nodes are stationary or
rather some are mobile.

Another classification is into the following two types:
{\em broadcast networks\/} and {\em point-to-point networks}.
The former are a number of nodes which share a communication channel;
a packet sent by any node is received by all the nodes.
A point-to-point network is a collection of nodes of which certain pairs 
are connected by transmission links; a packet sent from a node 
and destined at some specific node may need to hop through a sequence 
of nodes, along the links that connect them, until eventually it arrives 
at the destination.
Not unexpectedly, LANs are often broadcast networks and WANs usually
point-to-point.

Some of these taxonomies have little relevance to the principles governing
design of algorithms for communication protocols, on either the network or 
media-access levels. 
For instance, the fact that a packet comes over a copper wire link and
has covered a large distance is less relevant than what happens if
two packets arrive simultaneously at a node, because it may be the case
that both of them cannot be successfully received, and hence need to be sent
again.

We consider mainly synchronous networks, this means the following.
Each node has access to a clock. 
A clock cycle is called a {\em step\/} or a {\em slot}.
The clocks are assumed to start at the same moment, 
with possibly different initial clock values.
All the clocks tick simultaneously at the same rate.
This defines the {\em locally synchronous\/} model.
If, moreover, all the clocks show the same number, in other words, have access
to a {\em global clock}, then the model is {\em globally synchronous}.
We assume this stronger model, unless stated otherwise.

A {\em message\/} is a finite string of bits.
A {\em packet\/} is a message supplied with additional information 
to facilitate its traversal through a network.
A packet is assumed to have such a size that it can be transmitted between 
two nodes in one step.

A useful (partial) specification of a communication network is given by a 
directed graph, in which the nodes correspond to processing and 
communicating units, and the arcs denote ability to send messages directly.
If there is an arc from $v_1$ to $v_2$ then node $v_1$ is a {\em neighbor\/} 
of node~$v_2$, while $v_2$ is said to be {\em reachable\/} from $v_1$.
A node can send packets directly only to these nodes that are reachable
from it.
If communication in a network is over physical links then an arc corresponds 
to such a connection, and if nodes communicate by electromagnetic waves then 
all the nodes in the range of a node are reachable from it.

Additionally, we need to make clear how nodes
handle multiple messages concurrently.
In general, a node of a network may have the capability to 
send different messages to any subset of adjacent nodes in a step.
Similarly, a node may be either able to successfully receive all the incoming
messages in a step, or it may accommodate just a certain subset of them, 
the size of which may depend on the capacity of its buffer, or in an extreme 
case, if many packets come to a node in a step then none of them 
is successfully received.

The area of our considerations is radio networks.
What distinguishes them from other communication 
networks is the way nodes send and receive messages in a step.

\begin{myquote}
{\sf In a radio network:} 
If a node performs a send operation then it sends out just one message, 
and copies of this message are delivered by the next step to all 
the reachable nodes.
\end{myquote}

The mode of receiving information by the nodes of a radio network is restricted,
a node can successfully receive a message only if  exactly one was sent to it 
in a step.
If many neighbors send messages simultaneously to a node
then all of them are delivered to the recipient but they are
received as garbled.
If a message is received by node $v$ in its correct form then we say that $v$
can {\em hear\/} the message.

\begin{myquote}
{\sf In a radio network:}
A node can hear a message only if it was sent by its neighbor, and it was the
only neighbor that performed a send operation in that step.
\end{myquote}

It is convenient to assume that if a node does not hear a message then instead
it can hear some {\em noise}, which is distinct from any meaningful message.
If no message has been sent to a node then it hears the {\em background noise}.
If a node receives more than one message then we say that a {\em collision\/}
or a {\em conflict\/} occurred at the node.
If a collision happens then the node hears the {\em interference noise}.

Radio networks are categorized into four groups by the following two 
independent taxonomies.
If the nodes of a network can distinguish the background noise from the
interference noise then the network is said to be {\em with collision 
detection}, otherwise it is {\em without collision detection}.
The next categorization is with respect to the topology of the underlying
graph.
If this graph is complete bi-directional, that is, 
for each pair of nodes $v_1$ and $v_2$ there is an arc from $v_1$ to $v_2$, 
then the network is said to be {\em single-hop}, 
otherwise it is {\em multi-hop}.

Single-hop radio networks with collision detection are also known as 
{\em multiple access channels\/} or {\em broadcast channels}, 
and are a special case of broadcast networks.
A node is usually called a {\em station} in the context of single-hop 
radio networks.
Since a broadcast performed by a single station makes the message 
heard by all the stations, such a broadcast is called {\em successful}.
Multi-hop radio networks are also called {\em packet\/} 
or {\em point-to-point\/} ones.

Mobile users/nodes can form networks without any fixed infrastructure.
Then the nodes may have a limited knowledge about the current topology 
of the network.
We assume that the size~$n$ of the network is known, unless stated otherwise.
Moreover, the nodes are assumed to have been assigned unique identifying 
numbers (IDs), the range of IDs being $[1..N]$, where $N=\Theta(n)$. 
Each node knows its ID.
Distributed communication protocols are often robust enough 
to handle a situation when nodes do not know the IDs of their adjacent nodes.
If we discuss such protocols then the underlying network is said to 
be either {\em ad-hoc\/} or {\em unknown\/} or {\em of unknown topology}.
We assume that multi-hop radio networks are ad-hoc,  
when considering distributed communication protocols for them, 
unless stated otherwise.

A model related to radio networks is that of optical communication
(see~\cite{Gereb-GrausT92,GoldbergJLR97,GoldbergJM98}).
To explain the relationship, notice that a radio network can be interpreted 
as operating under just one wave frequency used by all the nodes.
In an {\em optical communication parallel computer\/} (OCPC) 
each node~$v$ is assigned its own frequency $F(v)$.
Node~$v$ can simultaneously receive any message placed on $F(v)$ (possibly
garbled) and send a single message to the channel of any other node.
Node~$v$ can {\em hear\/} a message delivered by $F(v)$ only if it 
was just a single message placed on it in a step.
We can see that OCPC is similar to a single-hop radio network in two 
respects: first, it can be interpreted as a communication network with the
topology of a complete bi-directional graph, and secondly, if many
messages are send to a node then it cannot hear any of them.
The main difference between the models of OCPC and single-hop radio-networks
is that the operation of sending in an OCPC network delivers the message to 
just a single recipient node rather than to all the stations.
A node~$v_1$ sending to channel~$F(v_2)$ cannot 
detect the collisions at~$F(v_2)$ but at~$F(v_1)$. 
If a message is successfully delivered in a single-hop radio network
then the sender also can hear the message, which serves as an acknowledgement. 
This mechanism can be implemented indirectly in the OCPC: let each packet carry 
both a message and the identification of its sender, then let after each 
``sending step'' another ``acknowledgement step'' follow, when all the nodes 
that have successfully received messages send confirmation messages to 
the senders.
Our discussion of optical communication protocols will be limited to 
situations in which there is some connection with  radio-network 
communication.

\section{Single-hop networks}

Most of the work done about randomized communication in radio networks 
has concerned multiple access channels.
In this section we first give a historical background, then discuss various
methodologies used to investigate problems in this area.

\subsection{History and existing systems}

The first communication network with the semantics 
of a (single-hop) radio network seems to be the system Alohanet~\cite{Abramson70,AbramsonK1973}.
developed in the beginning of the 1970's at the University of Hawaii.
The system used packet radio communication between the central computer 
and its terminals on the campuses on remote islands. 
That is how a ``broadcast channel'' was born. 
This historical development is a justification to classify as 
``radio communication'' all modes of communication in which 
conflicting packets cannot be heard, even if the electromagnetic waves are
not used as a medium to propagate signals.

The work on the Alohanet system has started already at the end of the 1960's.
As Abramson~\cite{Abramson-TIT85} put it, the key idea was: 
``not the use of radio communications for computers, but the use of 
a broadcast communication architecture for the radio channel.''
The developers of Alohanet realized that a protocol was needed to handle 
bursty traffic, with most of the stations being idle for most 
of the time, hence techniques like time-division multiplexing were 
not considered suitable.
The protocol invented for Alohanet is called Aloha.
Its most relevant feature is that control is distributed among the 
attached stations, and arbitration of access is statistical.
This is done in Aloha in an exceptionally simple way: the stations needing to
broadcast backlogged packets keep tossing independent coins to decide 
if to perform a broadcast in a step or rather to pause.
Actually Alohanet used the so called unslotted Aloha protocol, 
with the stations not assumed to be synchronized and making broadcast 
attempts at arbitrary times (see~\cite{BertsekasG92}).

The use of satellite and other channels implementing 
the multiple access channel followed in subsequent years.
Among them was Ethernet~\cite{MetcalfeB76} developed in the mid 1970's.
It is a communication system facilitating broadcast of data packets 
among locally distributed computers.
The medium of communication is usually a coaxial cable.
In Ethernet again the arbitration of conflicts is distributed and randomized,
the conflict-resolution protocol is the so called 
{\em truncated binary exponential backoff}.
It operates as follows: a new packet is transmitted immediately,
the time of the next attempt, after $k$ unsuccessful transmissions, 
is selected with a uniform distribution on $\{ 1,2,3,\ldots,2^{\min(10,k)}\}$, 
until $k=16$ when the packet is discarded.
Results of measurements of the performance of existing Ethernet 
installations can be found in~\cite{ShochH80}.
This paper reports results of an experiment that was carried out on 
$120$~stations;
very low latency and few collisions were experienced  under normal loads,  
the channel utilization approached~$98$\% under artificially generated 
very heavy load.
It is because of this remarkable performance that Ethernet is so popular.
The stability of the protocol is guaranteed by a possibility to discard 
packets, when the channel gets really jammed, so it is not a ``clean''
protocol from a theoretical point of view.

\subsection{Formal models}

Consider a single-hop radio network with some of the stations
storing packets that need to be broadcast on the channel.
The stations compete for access to the channel, and if randomized 
and decentralized control is used then conflicts for access are inevitable. 
Resolving such conflicts is the problem we want to be able to handle.
Distributed protocols used to provide such control and arbitrate among 
competing stations are called {\em conflict resolution protocols}, 
we will simply call them {\em protocols}.

A station is said to be {\em active\/} if it is ready to transmit a packet.
We assume that there is some external mechanism to generate messages and assign
them to stations with the purpose of broadcasting them on the channel.
The process of assigning new messages may be either performed in advance
or done on-line.
In the latter case, if a station does not hold a message in the beginning of
step~$i$ and it is assigned a message in step~$i$ then we say that the station
was {\em activated in step~$i$}.
In the case when new messages are allocated to stations on-line some 
confusion with terminology may occur.
Namely, if new messages are assigned dynamically to stations then
we say that they {\em arrive\/} to stations, which is consistent
with the term ``arrival rate.''
On the other hand, the nodes of a radio network communicate among themselves
by sending packets containing messages, the incoming packets also are 
acquired by nodes, but in such a situation we say that they are 
{\em received}. 
We hope that  ``received packets'' will not be confused with 
``arriving messages.''
A message/packet that has been attempted to be broadcast in vain
is called a {\em backlogged\/} one.
A message which in step $t-1$ became the one that a station will attempt to
broadcast starting from step~$t$ is a {\em new\/} message at step~$t$.

The problem of conflict resolution may be investigated
either in a static or a dynamic scenario.
In a former one, all the messages are created and allocated to the 
stations prior to the beginning of a protocol. 
Additionally, a station is allocated either just 
a single message or no messages at all.
In a dynamic scenario, new messages keep being generated all the time and 
assigned to stations, which then try to broadcast them on the channel.
Independently, we may have either an infinite or a finite number of users,
depending on whether the set of stations is potentially unbounded or finite, 
respectively.
Having an infinite collection of stations may seem to be strange at first, 
but the motivation is to treat messages as running protocols 
themselves, independently from the other messages. 
We distinguish the following four settings.

In a {\em static finitely-many users model\/} there are a total of 
$n$~stations, some $k$~stations among them are allocated  messages prior 
to the start of a protocol.
Each of the numbers~$k$ and~$n$ may either be known or not to the 
stations.

In a {\em static infinitely-many users model\/} there are countably 
infinitely many stations. 
Each of them performs a Bernoulli trial with some probability $p$ 
of success prior to the start of a protocol.
A success results in a message placed at the station, 
whereas a failure means that there is no message located there and never will.
The probability~$p$ is referred to as {\em the probability to hold a 
message}.

A {\em dynamic finitely-many users model\/} is specified, first, 
by the number $n$ of stations, and secondly, by the mode of arrivals 
of messages.
Usually the arrival process is Bernoulli, that is, in each step at each 
station a message is generated with some constant probability~$p$, 
independently of the other stations and steps.
Each station has a buffer of potentially unbounded capacity, to store
backlogged messages.
It operates as a queue, that is, on the first-come-first-served basis.
Otherwise, it it operated on the first-come-last-served basis, 
then some protocols in which each message is eventually dispatched 
successfully with a queue would deteriorate in such a way that 
the probability that a message is eventually sent is less than one.
The current size of the queue of the $i$-th station is usually denoted 
as~$q_i$.

In a {\em dynamic infinitely-many users model\/} the number of stations is
again countably infinite.
In each step a number of new messages are generated. 
Each of the messages is allocated to a new station.
A station is identified by the time it obtained a message to transmit.
The number of messages generated in a step has a fixed distribution,
exactly~$k$ new messages are generated in a step with some probability 
$\lambda_k$.
If it is not stated otherwise then the distribution is Poisson with 
parameter $\lambda$, that is, 
$\lambda_k=e^{-\lambda}\lambda^k/k!$ (throughout this article~$e$
is the base of the natural logarithm).

Variations on the theme of models are possible.
For instance, a static finitely-many users model may be defined by 
stipulating that some~$N$ messages have arrived, with the arrival times being
real numbers in the interval $(0,1)$ generated by a Poisson process,
where simultaneously there are $N$ stations, each station learns its 
corresponding message at its arrival time (see~\cite{Pippenger81}).

For the static finitely-many users model two tasks are of interest.
First is to make some station successfully transmit its packet, 
which is the {\em static selection problem}.
Second is to make all the $k$~stations with packets transmit successfully,
which is the {\em static all-broadcast problem}.
For the static infinitely-many users model, the task is to make each 
message broadcast successfully with probability one, 
simultaneously achieving the maximum transmission rate.

For the dynamic case, the basic question is whether a given protocol is 
stable, in the sense that it is practically operational, without the channel
eventually getting clogged.
We associate a Markov chain with  each of the dynamic models.
In the finitely-many users case, a state description includes the sequence 
of numbers of packets waiting at the stations.
In the infinitely-many users model, a state includes the total 
number of backlogged packets.
Stability of protocols is usually expressed in terms of 
properties of the underlying Markov chain.

One might expect differences in performance of similar protocols if they are
analyzed in both the infinite and finite-numbers-of-users models.
Observe that with infinitely many stations a newly generated message may be 
immediately broadcast on the channel, what contributes to burstiness 
of the traffic and hence is destabilizing.
With finitely many stations, a message is first 
put into a buffer of the station to wait till its predecessors in the buffer 
have been handled, what stabilizes the traffic.

\subsection{Conflict resolution }

First let us discuss a certain basic feature of protocols, which is 
related to synchrony.
A protocol is {\em slotted\/} if attempts to broadcast are 
always performed at the beginnings of slots.
We will consider only slotted protocols,
mainly to simplify the analysis.
They need at least a local synchronization, which is an additional requirement 
on a network, so we may ask a question if it can be harnessed to 
improve performance of protocols.
This indeed is the case, the following is a heuristic explanation.
Suppose that we have a certain unslotted protocol $\cal UP$.
Consider the slotted protocol $\cal SP$ obtained from $\cal UP$ by using
the following more stringent regime to send packets: 
a packet to be broadcast during a slot is forced to be broadcast 
exactly at the first coming border between two slots.
Let a slot last some $t$~time units, and suppose that it takes the full 
slot size  to perform the operation of transmitting a packet.
A collision of packets sent by $\cal SP$ happens if they were
sent during the same $t$ time units constituting a slot.
On the other hand, two packets processed by $\cal UP$ collide if they 
were sent in such a way that the time period from the start of the
transmission of the earlier one till the completion of transmission of the
later one comprise less than $2t$ consecutive time units.
Hence one may expect that the slotted protocol $\cal SP$ has twice 
as high a transmission rate as \,$\cal UP$ (see~\cite{KellyM87}).

A protocol may grant transmission rights to a number of stations
at the beginning of a step, we say then that the protocol {\em enables\/} 
these stations.
Each enabled station transmits a message, provided it is active.
Protocols may refer to stations in various ways, depending on the formal
model in case.
In a static model, each station knows its unique identification number (ID), 
and protocols enable stations by referring to these IDs.
In a dynamic model, stations are not identifiable by unique IDs,
and protocols refer to them by their activation time. 
In a static infinitely-many users model a protocol is assumed to know the
probability to hold a message by a station.
In a dynamic case, the knowledge of the rate of arrival of messages 
is usually not needed by protocols.
This is because if the arrival rate is smaller than the maximum stable
transmission rate of a protocol then the actual transmission rate 
automatically adjusts to the arrival rate.

Packets carry messages which bring some useful information to the stations.
Our philosophy is to work with ``clean'' protocols in which 
this information does not affect the future schedule of broadcasts.
This also simplifies implementation of protocols.
In particular, we do not want to consider any reservation mechanism,
the reasons for this are as follows. 
A static reservation system, like time division multiplexing, would assign
slots to stations on a fixed basis, what would be a waste of time if the
traffic is bursty.
A dynamic reservation system, say when a packet carries a message about
reservation of some of the following slots, may make single stations 
dominate the channel for long periods of time.

The more restricted a protocol is, in terms of the information retrieved
from the channel that is uses in scheduling future broadcasts, 
the more viable and versatile it is.
The categorization of protocols is usually made based on the kind of feedback 
they receive and use.
Consider first the single-hop radio network without collision detection.
Then two kinds of feedback in a step are possible to be received by every
station:

\begin{myquote}
\begin{description}
\item[\tt meaningful message:] a successful broadcast;
\item[\tt silent channel:] either no attempt of broadcast has 
been made or a conflict among multiple attempts occurred.
\end{description}
\end{myquote}

If a station is not making an attempt to broadcast in a step and it hears a
silent channel then it gains no information.
This is because it is our assumption that packets have the same size 
and it takes a full slot to broadcast a packet, 
hence a busy slot following a silent one does not mean that processing of 
a packet will be continued and the channel will be busy.
A station attempting a broadcast can learn if it was successful by listening to
the channel: it can either hear its own message (a success) or silence (a
collision).

Protocols that schedule broadcasts of the current packet 
based only on the history of the previous attempts to broadcast it
are called {\em acknowledgement based}. 
The explanation for this term is that each station relies only on the 
acknowledgement of a successful broadcast of its own packet.
Such protocols may operate in both the channel with collision detection and 
without it, since a station performing a broadcast can always 
detect a collision.
An acknowledgement based protocol refers to the following 
restricted history, which is a sequence $\langle t_1, t_2, t_3,\dots,
t_k\rangle$, with $t_i$ being the number of step of the $i$-th unsuccessful 
attempt to broadcast the current packet, among $k$ unsuccessful 
attempts made so far. 

In a channel with collision detection every station may have 
the following  feedback:

\begin{myquote}
\begin{description}
\item[\tt  meaningful message:] a successful broadcast; 
\item[\tt  silent channel:]  no attempt of broadcast has been made; 
\item[\tt interference noise:] a collision caused by many simultaneous
attempts to broadcast.
\end{description}
\end{myquote}

Because of these three possibilities such channels are often called
{\em ternary feedback}.
For the channel with collision detection, it may be useful if every station 
listens to the channel all the time and remembers the events that happened.
Usually the channel history is stored as a sequence 
$\langle u_1,u_2,u_3,\dots\rangle$, where value $u_i=0$ means a silent 
channel with no attempt to broadcast during the $i$-th step,
value $u_i=1$ corresponds to a successful broadcast during step~$i$ by some 
station, and finally value $u_i=*$ means a collision in step~$i$.
Additionally, a station might remember its own attempts to broadcast.
Protocols that rely on such a history are called {\em full sensing}.

We may also consider scenarios when stations obtain more information
from the channel.
If each station gets to know the number of stations involved in a collision 
in the current step then the channel and protocols are said 
to be {\em with collision-size detection} or {\em with control}.

Let us consider a dynamic scenario.
A protocol is said to be {\em stable\/} if the associated Markov 
chain is ergodic, that is, irreducible, positive recurrent and
aperiodic~(see~\cite{Feller-vol1}).
Usually natural protocols are both irreducible and aperiodic, so only
positive recurrence is an issue.
The Markov chain of a stable protocol converges with time to a stationary 
distribution of states, each probability of arriving at a state 
being the inverse of its expected return time.
In particular, the expected time to wait for the backlog of packets
to disappear is finite, as is the time to wait for a given message in the
system to be sent successfully.

A related issue is that of buffer size.
One can construct an example of a Markov chain with the states 
$\{0,1,2,\dots \}$ which is ergodic and such that the average state number for
the stationary distribution is infinite.
Interpreting the size of buffer as the state number we can see that a stable
protocol may have an infinite expected buffer size,  with respect to the
stationary distribution.
The time spent by a message waiting depends on the size of the
buffer at the step when the message was assigned to the station.
Both this dependence and the stationary distribution determine the expected
waiting time of a newly generated message.
If the expected buffer size is infinite then such is the expected waiting time.
Hence if a protocol has a finite average waiting time then this is a 
stronger property than stability, we call it {\em strong stability}.

If a protocol is not stable then the associated Markov chain can be either 
null recurrent or transient.
The former means that with probability one the backlog will become empty 
eventually but the expected time for this to happen is infinite.
The latter means that with probability greater than zero the system will 
eventually be jammed, in the sense that the backlog 
will never be empty.

It is possible that each message is eventually delivered but the protocol 
is not stable.
For instance, consider the example of a finite number of users and a stable 
protocol~$\cal SP$. 
Convert $\cal SP$ to $\cal UP$ by the following modification:
if a station has just a single message then it stops running $\cal SP$
temporarily and does not attempt to broadcast the message until a new 
message is generated, then it resumes $\cal SP$.
This protocol is unstable in the sense that the state of empty
backlog never happens after a finite number of steps, but the expected 
time to broadcast each message is finite.
Needless to say, protocols are designed so that such 
a strange behavior does not happen.
Clearly the problem with protocol $\cal UP$ is that the associated Markov 
chain is not irreducible.

A protocol that is not stable may still be considered good if one of
the following properties hold: 

\begin{myquote}
\begin{description}
\item[\rm (a)] 
The expected delay between two consecutive messages sent is finite.
\item[\rm (b)] 
Each message is sent eventually with probability one.
\item[\rm (c)] 
The number of messages sent successfully is infinite with 
probability one.
\end{description}
\end{myquote}

We say that a protocol is {\em weakly stable\/} if it has property~(c) above,
which we consider to be the weakest satisfactory behavior of a protocol.

A protocol is {\em first-come first-served\/} (FCFS)
if for any two messages $m_1$ and $m_2$, such that $m_1$ is generated at step 
$t_1$ and successfully broadcast at step $t_1'$, while $m_2$ is generated 
at step~$t_2$ and broadcast at step~$t_2'$, we have that if 
$t_1\le t_2$ then $t_1'\le t_2'$.
For FCFS protocols, property~(a) implies weak stability, and property~(b) 
is equivalent to it.

In a dynamic model, during each step $i$ some $A(i)$ new messages arrive.
If the limit  $\lambda=\lim_{t\rightarrow\infty}\frac{1}{t}\sum_{i\le t}A(i)$ 
exists then $\lambda$ is called the (average) {\em arrival rate};
throughout this chapter the letter $\lambda$ will denote 
the arrival rate, unless stated otherwise.
In the infinitely-many users model, if new messages are generated
with the Poisson distribution with parameter~$\lambda$, or 
similarly in the finitely-many users model of $n$ stations,
if each station generates a new message with probability $\lambda/n$
independently over the steps and the stations, 
then this $\lambda$ is the average arrival rate. 
Similarly, suppose some protocol $\cal P$ governing the operation of a 
channel is used, and at step $i$ some $Z(i)$ packets are successfully broadcast
over the channel, where $Z(i)$ is a random variable equal to either zero or 
one.
If the limit  $\sigma=\sigma_{\cal P}=\lim_{t\rightarrow\infty}\frac{1}{t}\sum_{i\le t}Z(i)$ 
exists then $\sigma$ is called the (average) {\em transmission rate\/} or
{\em departure rate\/} of protocol~$\cal P$.
The ratio $\sigma/\lambda$ is a measure of efficiency of the protocol
and is called its {\em throughput}.
Clearly a throughput is at most one.
It is necessary for a protocol to attain a throughput of exactly one 
if it is to perform well in terms of stability,
because otherwise the backlog increases and the channel is eventually clogged.
On the other hand, it is possible that both the arrival and departure rates 
are one and the protocol is not stable (\cite{HastadLR-SICOMP96}): consider a single station,
which always broadcasts a message, provided it has one, and such that 
during each step either two new messages arrive or none, each such event
happening with probability~$1/2$ independently over the steps.
This system behaves like a symmetric random walk on a line, hence it is null
recurrent.

The throughput is a relative departure rate, that is, the transmission rate 
related to the arrival rate.
For a static model, there is no notion of time regarding 
the generation of new messages, and hence no arrival rate. 
However, for the static infinitely-many users model, the transmission rate 
has the same intuitive meaning as the throughput in the dynamic case.

The throughput of a protocol is an incidental measure of the 
utilization of a channel, because it may happen that the same protocol 
has either the maximum unit throughput or a throughput smaller than one, 
depending on the magnitude of the arrival rate~$\lambda$. 
It would be good to have the ``ultimate performance'' measure of a channel,
indicating the limits of its performance in absolute terms.
Such a measure will be called capacity.
We assume the infinitely-many users model, otherwise this 
measure becomes parametrized by the number of stations.

In a static model, where each station holds a message with some 
probability~$p$, the performance of a protocol ${\cal Q}$ is measured by its 
departure rate.
Notice that it is easy to attain a transmission rate equal to $p$ 
by trying all the stations one by one; 
then the transmission rate approaches one if $p$ approaches one.
This indicates that the challenge for a protocol is when the probability 
to hold a message is small rather than large, and is a justification of the
following definition of capacity.
Let $\cal Q$ be a protocol for a static model. 
Let ${\sigma}_{\cal Q}(p)$ be the departure rate 
of the protocol $\cal Q$, with $p$ being the probability to hold a message; 
recall that we assume that a protocol knows the probability~$p$.
We define the {\em tenacity\/} of protocol~$\cal Q$, denoted~TN$(\cal Q)$,
to be the limit $\lim_{p\rightarrow 0} {\sigma}_{\cal Q}(p)$ if it exists.
The {\em capacity\/} of a channel is defined to be 
the supremum of the tenacities~${\rm TN}({\cal Q})$ over all  
protocols~$\cal Q$.

The capacity of a channel in a dynamic case can be defined similarly.
Let $\cal R$ be a protocol. 
Suppose new messages are generated with a Poisson distribution with 
the arrival rate~$\lambda$.
Let ${\sigma}_{\cal R}(\lambda)$ be the transmission rate 
of protocol $\cal R$ for this model of arrivals. 
Define the {\em tenacity\/} of protocol $\cal R$, denoted again as 
TN$({\cal R})$, to be the supremum of the arrival rates~$\lambda$ for 
which the throughput of~${\cal R}(\lambda)$ is also~$\lambda$.
The capacity of a channel is now defined to be the supremum of tenacities
TN$({\cal R})$, over all the protocols $\cal R$.
We might say briefly that, in the dynamic case,
the capacity of a channel is its maximum stable transmission rate.
The following is an alternative definition of capacity:
it is the supremum of the throughputs $\sigma_{\cal R}(1)$ attained 
for the Poisson arrivals with the unit rate, over all the protocols~$\cal R$.

There is a relationship between Bernoulli distributions of messages for the
static model and Poisson distributions of new arrivals for the dynamic model.
The Bernoulli distribution of the number of successes in $n$ trials, each with
the probability~$p$ of success, converges to the Poisson distribution with 
parameter~$\lambda$ if $n\rightarrow\infty$ and $p\cdot n=\lambda$ 
(see~\cite{Feller-vol1}).
Hence when the probability~$p$ to hold a message converges to zero, 
we may model the distribution of messages in the static model, in the limit,
by a dynamic model with a Poisson distribution.

Sometimes it is useful to interpret the messages waiting at stations as 
customers waiting in queues, to apply the methods of queuing 
theory~(see~\cite{Kleinrock-book75-76}).
Suppose $L(t)$ is the number of customers at step~$t$.
The limit $L=\lim_{t\rightarrow\infty} L(t)/t$ is the 
{\em average number of customers}.
If a message $m$ arrives at step $t_1$ and is successfully broadcast
at step $t_2$ then $t_2-t_1$ is its {\em waiting time}.
Let $W_t$ be the average waiting time of the messages that have been sent 
through the channel by step~$t$.
If the limit $W=\lim_{t\rightarrow\infty}W_t$ exists then it is called the
{\em average waiting time\/} of a customer.
If any two among the parameters $W$, $L$ and $\lambda$ are finite
then the third one is finite too, and they satisfy the Little's formula 
$L=\lambda\cdot W$ (see~\cite{Stidham-OR74}).
Hence if a protocol behaves nicely then we do not need to study separately 
$L$ and $W$ for a given arrival rate, estimates of one of them immediately
yield estimates for the other one.

Many protocols, especially acknowledgement-based ones, operate by making 
each active station broadcast a packet at step $t$ with some 
probability~$f_t$, which depends on the history of feedback from the channel,
independently of the other steps and stations.
An especially simple scheme of such an algorithm is determined by a fixed 
sequence of probabilities $\langle h_0, h_1, h_2, h_3,\ldots\rangle$
as follows. 
When a new packet needs to be handled, in each step a coin is tossed 
with probability~$h_0$ of obtaining heads, until the first heads come up, 
then the packet is broadcast immediately;  usually it is assumed that $h_0=1$.
In case of a failure, a new sequence of coin tosses is performed, 
a toss per step, each with the probability~$h_1$ of obtaining heads, 
a broadcast is performed when heads come up.
This is continued through a sequence of failed broadcast attempts, using
a coin with probability $h_i$ after the $i$-th failed attempt, 
until eventually the packet is successfully broadcast.
So the protocol backoffs from the probability $h_{i-1}$ after the $i$-th 
failure and tries the next probability~$h_i$,
that is why such protocols are called {\em backoff}.
We may also call them {\em randomly oblivious}, because the sequence of
probabilities $\langle h_i \rangle$ is fixed in advance.
The number of failed attempts of station~$i$ to broadcast the current packet 
is maintained in the {\em backoff counter}, denoted~$b_i$.
The probabilities $\langle h_i \rangle$ are called {\em backoff probabilities}. 
The function assigning the probability~$h_i$ to number~$i$ is referred to as
{\em backoff rate}.
Backoff protocols are interpreted naturally as Markov chains, by defining the
states to be either sequences of the backoff counters $b_i$ of the stations 
with backlogged packets, in the case of infinitely-many users model, or vectors 
$\langle b_1,\ldots,b_n,q_1,\ldots,q_n\rangle$ of the backoff counters $b_i$
together with the queue sizes $q_i$ in the case of 
finitely-many users model of $n$~stations.

Certain full-sensing algorithms operate in a similar fashion as randomly
oblivious ones and use coin tosses to decide whether to attempt 
a broadcast, but they modify the probability $f_t$ of heads coming up
in step~$t$ on-line, by a function of the history of the ternary feedback 
of the channel up to step~$t$; we call such algorithms {\em randomly adaptive}.
Such randomized algorithms for the channel with collision-size detection
have this probability~$f_t=f_t(k)$ depending also on the number~$k$ 
of stations colliding in step~$t-1$.

A protocol called {\em Aloha\/} is the simplest randomly oblivious one, 
it operates as follows.
A new packet is broadcast immediately.
A backlogged packet is transmitted repeatedly in the following steps until 
success, each attempt made independently with some constant probability~$f$.
{\em Aloha with control\/} is a class of protocols in which each station with a
message transmits with the probability equal to about $1/k$, where~$k$ is the
current number of stations with messages.
Such protocols need to be further specified, since the strongest feedback 
from a channel that we consider provides merely the number of 
colliding packets, if any.
A backoff protocol with $h_i=\Theta(a^{-i})$, for $a>1$, is called an 
{\em exponential backoff}, the number $a$ is called its {\em factor}, 
in particular if the factor is $a=2$ then the resulting protocol is 
{\em the binary exponential backoff}.
A backoff protocol with $h_i=\Theta(i^{-b})$, for a constant $b>0$, is called 
{\em polynomial}, which in turn is {\em sublinear\/} if $b<1$, 
{\em linear\/} if $b=1$, and finally {\em superlinear\/} if $b>1$.

An alternative approach to define  backoff protocols is by specifying 
an increasing sequence $\langle s_0=1,s_1,s_2,\ldots\rangle$ of integers. 
A station with a new message acquired at step $t$ makes a sequence of 
attempts $\langle t_0,t_1,t_2,\dots\rangle$ to broadcast it, 
until the first success. 
First $t_0=t+s_0$, then $t_1$ is a random element of the interval 
$[t+s_0+1,t+s_1]$, and in general the time step $t_i$, for $i>0$, 
is randomly selected from the interval $[t+s_{i-1}+1,t+s_i]$.
This is similar to how the protocol of Ethernet operates,
which uses the sequence $\langle 2^i\rangle$, for $i=0,1,2,\ldots$.
Consider both such a protocol determined by the sequence 
$\langle s_0=1,s_1,s_2,\ldots\rangle$, and also the one determined by the
probabilities $\langle h_i \rangle$, where $h_0=1$ and 
$h_i=1/(s_i-s_{i-1})$, for $i>0$. 
We expect that they are equivalent, as far as 
properties like stability are concerned.
Backoff protocols defined by a sequence of time intervals may be 
called {\em unmodified}, because historically the first backoff protocol 
was that used by Ethernet, which is determined by intervals.
Accordingly, a backoff protocol defined by a sequence 
$\langle h_i \rangle$ of probabilities is called {\em modified}.

\section{Classifying Markov chains}
\label{tools}

To answer questions about stability of protocols, we need criteria
to answer questions about either positive recurrence or null recurrence or
transience of a given Markov chain.

Let $\langle X_i\rangle$ be a Markov chain with the set of positive integers
as the set of states.
Let $\langle P_{ij}\rangle$ be the transition matrix, that is,
$\Pr (X_{k+1}=j\mid X_k=i)=P_{ij}$.
The chain is recurrent if for each $i$ the event $X_n=i$ occurs infinitely
often with probability one.
The chain is transient if it diverges to infinity with probability one.
In what follows we will consider only irreducible and aperiodic Markov chains,
they will also have a countable state space.

Sufficient conditions for the ergodicity and
recurrence of (irreducible and aperiodic) Markov chains 
were given by Foster~\cite{Foster53}.
Various extensions have been found later, a very simple one is by  
Pakes~\cite{Pakes69}.
Let $\langle X_n\rangle$ be a Markov chain, with states being the nonnegative
integers.
Let the {\em drift at state~$i$} be the quantity 
\[
{\cal D}(i)=\mE \, [X_n-X_{n-1}\mid X_{n-1}=i]\ .
\]
The Pakes criteria, for an irreducible and aperiodic Markov chain,
are as follows: 
\begin{enumerate}
\item
If the drift is finite for all states, and bounded above by a negative 
constant for all but finitely many states, then the chain is ergodic.
\item
If the drift is bounded above by zero, for all but finitely many states, 
then the chain is recurrent.
\end{enumerate}

Various criteria to classify Markov chains have been formulated in terms 
of a positive real function~$F$ defined on the set of states.  
Such functions are called either {\em test\/}
or {\em Lyapunov\/} or {\em potential\/} ones.
For instance, the test functions used in~\cite{HastadLR-SICOMP96}, to investigate 
backoff protocols for the finitely-many users model, are of the form
\[
F(\langle q_1,\dots,q_n,b_1,\ldots,b_n\rangle)
=
H_1(n)\cdot\sum_{i=1}^n q_i + \sum_{i=1}^n H_2(b_i)+H_3(n)\ ,
\]
for certain functions $H_1$, $H_2$, and $H_3$, where numbers~$q_i$ are  
queue sizes and numbers~$b_i$ are backoff counters.

Another example of a useful criterion is the one which states that,
for a given  Markov chain and test function $F\ge 1$, the following are
equivalent:
\begin{enumerate}
\item
The chain is ergodic with the stationary probability $\pi$ and a finite 
expected value of $F$ with respect to $\pi$. 
\item
There is a finite set of states $C$ such that the expected values of the sums
\begin{equation}
\label{eq1}
\sum_{t=1}^{T(C)} F(s_t)
\end{equation}
are finite for all $s\in C$, if the chain starts at $s$ and proceeds 
through the states $s_1,\ldots,s_{T(C)}$, where  $T(C)$ is the time needed 
for the chain to return to $C$ after the start.
\end{enumerate}
It was used in~\cite{HastadLR-SICOMP96}, see~\cite{MeynT09-book} for a proof.

Given a test function~$F$, the drift can be generalized to {\em $F$-drift},
which  at state~$i$ is the quantity 
\[
{\cal D}_F(i) = \mE \, [F(X_n)-F(X_{n-1})\mid X_{n-1}=i]\ .
\]
The Foster's criterion generalized with such a drift is as follows:

\begin{myquote}
A Markov chain is ergodic if and only if there exists a positive real
function~$F$ and a finite set of states~$C$ such that: 
\begin{description}
\item[\rm (a)] 
the drifts ${\cal D}_F(i)$ are uniformly bounded above by a 
negative constant, for each state~$i$ outside $C$, and 
\item[\rm (b)] 
the expectancies $\mE \,[F(X_{n+1})\mid X_n=i]$ are 
all finite for each state~$i$ in $C$.
\end{description}
\end{myquote}

This criterion gives some flexibility by allowing the drift to be calculated
with respect to a test function, but still it requires that its value 
decreases in just one step.
This requirement may be relaxed by allowing the decrease to happen in a
number of steps, which is bounded by another function.
Given two functions $F$ an $G$, where $G$ is integer valued, 
define drift by the formula:
\[
{\cal D}_{F,G}(i)
=\mE \, [F(X_{n+G(X_n)})-F(X_n)\mid X_{n}=i] \ .
\]
The following is a further generalization of the criterion:

\begin{myquote}
A Markov chain is ergodic if and only if there are two positive functions $F$
and $G$, where $G$ is integer-valued, a positive constant $\epsilon$, and a
finite set of states~$A$ such that the following holds: 
\begin{enumerate}
\item
The inequalities ${\cal D}_{F,G}(i) \le - \epsilon\, G(i)$ 
hold for all states~$i$ not in set~$A$. 
\item
The quantity $\mE \,[F(X_{n+G(X_n)})\mid X_n=i]$ is finite for all states~$i$ in 
set~$A$.
\end{enumerate}
\end{myquote}

This criterion was used in~\cite{Al-AmmalGM-TCSy01} to prove stability of the binary
exponential backoff for certain arrival rates in the finitely-many users 
model.
For proofs of these generalizations of the Foster's criterion 
see~\cite{FayolleMM1995-book}.

Kaplan~\cite{Kaplan79} showed that if a Markov chain has the following two
properties, then it is {\em not\/} ergodic:
\begin{enumerate}
\item
The drift is positive, for all but finitely many states.
\item
There are constants $B\ge 0$ and $0\le c\le 1$ such that the following
inequality 
\[
z^i-\sum_{j}P_{ij}z^j\ge -B(1-z) 
\]
holds for $z\in[c,1]$ and all but finitely many states~$i$.
\end{enumerate}
Property~2 above holds if the downward transitions are uniformly 
bounded,  that is,  $P_{ij}=0$ whenever $j<i-k$, for some constant $k>0$.  
For instance, let $X_n=X_{n-1}+a_n-b_n$, for~$n\ge 1$, be the number of 
packets in the system, with $a_n$~arrivals and $b_n$~packets
successfully dispatched in step~$n$. 
If $b_n$ is either $0$ or $1$, like in the multiple access channel, then 
the downward transitions are uniformly bounded.
The system is unstable if 
\[
\mE \, [ a_n\mid X_{n-1}=i ] > \mE \, [ b_n\mid X_{n-1}=i ] \ ,
\]
for all but finitely many $i$.

These criteria have later been extended in~\cite{Sennott87,SennottHT83,Szpankowski85,Szpankowski88,SzpankowskiR88}.

A criterion to prove that a Markov chain is transient was given by 
Rosenkrantz and Towsley~\cite{RosenkrantzT83}.
It is as follows:

\begin{myquote}\rm
Suppose the Markov chain $\langle X_i \rangle$ is irreducible and aperiodic, 
and that a nonconstant positive test function~$F$ exists such that 
the inequalities
\[
\mE \, [F(X_{i+1})\mid X_i=k] \le F(k)
\]
hold for every state~$k$. 
Then the Markov chain $\langle X_i \rangle$ is transient.
\end{myquote}

A proof can be based on the martingale convergence theorem.
Namely observe that by the assumption in the criterion the sequence
$F(X_i)$ is a nonnegative supermartingale with respect to 
$\langle X_i\rangle$.
Suppose, to the contrary, that the chain $\langle X_i\rangle$ is recurrent.
Take any two states $j_0$ and~$j_1$ such that $F(j_0)\neq F(j_1)$.
It follows from the recurrence that the events $X_i=j_0$ and $X_i=j_1$
occur infinitely often with probability one, and hence that
$F(X_i)=F(j_0)$ and $F(X_i)=F(j_1)$ infinitely often with probability one.
It follows that the limit of $F(X_i)$ does not exist, what contradicts the
martingale convergence theorem (see~\cite{Doob-book}).

A systematic exposition of related topics can be found in books by 
Fayolle, Malyshev and Menshikov~\cite{FayolleMM1995-book}
and Meyn and Tweedie~\cite{MeynT09-book}.

\section{Infinitely-many  users models}

In this section we consider single-hop networks, and concentrate on
the setting in which each newly generated message receives a dedicated
station, which then disappears after the message has been handled successfully.
This is a natural model to study the issue of channel capacity.
We present upper bounds, which reflect the inherent limitation on a
randomized conflict resolution, and lower bounds, provided by efficient 
algorithms.

\subsection{Upper bounds on capacity}

We assume that the channel gives a ternary feedback, unless stated otherwise.
The first upper bound on the capacity of the multiple access channel, 
for the infinitely-many users model with  Poisson arrivals, was
shown to be at most~$0.774$ by Pippinger~\cite{Pippenger81}
by an information theoretic argument.
Improvements of this estimate were given in~\cite{CruzH82,MikhailovT81,Molle82,TsybakovL87}.
The best known upper bound appears to be $0.568$, 
it was proved by Tsybakov and Likhanov~\cite{TsybakovL87}.
Some specialized bounds are known, if the class of protocols is restricted.
Panwar, Towsley and Wolf~\cite{PanwarTW85} showed that $0.5$~is an upper bound 
for the class of FCFS algorithms.
Goldberg, Jerrum, Kannan, and Paterson~\cite{GoldbergJKP04} showed that every 
backoff protocol is transient if the arrival rate is at least~$0.42$, and
for every acknowledgement-based protocol the same holds if
the arrival rate is greater than~$0.531$.

We sketch a reasoning given by Molle~\cite{Molle82} to estimate the capacity.
Consider the static infinitely-many users model
with the probability~$p$ to hold a message.
The trick is to analyze a conceptual algorithm which resorts to a helpful
genie.
The algorithm is in phases, a new one starts 
when the algorithm enables some new~$N$ stations. 
If a collision happens then the genie reveals the two smallest stations with 
packets involved in the collision, which are then enabled in the 
two following steps.
We use the following notation: $S_N$ is the probability of success, 
$I_N$ of an idle slot, $C_N=1-S_N-I_N$ of a collision. 
The transmission rate conditioned on~$N$ is
\begin{equation}
\label{eq2}
\rho_N=\frac{S_N+2C_N}{I_N+S_N+3C_N} .
\end{equation}
An optimal number~$N(p)$ of points to be enabled simultaneously at 
the beginning of a new phase is a nonincreasing function of~$p$.
If~$p$ is sufficiently large then enabling just a single point is optimal,
and the departure rate  is~$p$. 
Comparing this with the throughput conditioned on~$N$ we obtain that 
this happens for~$p>1/\sqrt{2}$.
Similarly, enabling two points, that is~$N=2$, is optimal for the 
probability~$p$ between $1/\sqrt{2}$ and the solution to the equation 
$\rho_2=\rho_3$, which is~$p\approx 0.568$.
Suppose we go to larger~$N$ and resort to help from the genie.
We may model Poisson arrivals by assuming $p\rightarrow 0$ while $N$ 
is chosen such that $p\cdot N$ converges to a constant value~$\lambda$.
Then the transmission rate given  by equation~(\ref{eq2}) takes the form
\[
\frac{2-(2+\lambda)e^{-\lambda}}{3-2(1+\lambda)e^{-\lambda}} \ ,
\]
which is maximized at $\lambda\approx 2.89$, yielding the departure rate
approximately equal to~$0.6731$.
This is an upper bound on the capacity of the multiple access channel.

\subsection{Algorithms}

Conflict resolution protocols process the feedback from the channel
and schedule the future broadcasts accordingly.
We consider three possible kinds of feedback: that giving the size of
the set of stations in conflict, the ternary feedback, 
and finally that of a single-hop network without collision detection. 

\subsubsection{Protocols with control}

Let us start the discussion of protocols from those that can rely on the 
most informative feedback: if a collision occurs then every station 
receives the number of colliding packets sent to the channel in the step.
This follows historical developments, because the study of properties 
of conflict-resolution protocols was started with investigating Aloha 
with control.

\Paragraph{Dynamic model.} 

Aloha with control is strongly stable for the arrival rate 
smaller than $1/e\approx 0.368$ (see~\cite{FayolleGL77,Ferguson75,Kleinrock-book75-76}). 
To show that this is an ergodic process, one can proceed as follows.
Let $n(t)$ be the number of active stations at step~$t$.
Suppose first that each active station knows the number~$n(t)$ 
at the beginning of step~$t$ and enables itself with the 
probability~$f(n(t))$.
The probability that exactly one active station will enable itself is 
\[ 
n(t)\cdot f(n(t))\cdot  (1-f(n(t))^{n(t)-1}\ .
\]
This is maximized for $f(n(t))=n(t)^{-1}$ and
the probability of a successful broadcast becomes 
\[
\Bigl(1-\frac{1}{n(t)}\Bigr)^{n(t)-1}\approx \frac{1}{e}\ .
\]
The sequence $\langle n(i) \rangle$ is a Markov chain.
If $Y_i$ is the number of arrived messages and $Z_i$ the number of 
successfully broadcast packets at step~$i$ then 
the chain is ergodic for $\lambda=\mE \, Y_i<\mE \, Z_i=1/e$, 
by the Pakes' criterion (see Section~\ref{tools}).

Now consider the real situation when the feedback at step~$t$ does not 
provide the number of active stations $n(t)$ but merely the number 
of the colliding packets at step~$t$.
We consider the following specific protocol.
The slots are partitioned conceptually into dynamic contiguous blocks.
In the first slot of a block, that is, just after all the stations trying to 
broadcast in the previous block have succeeded, all the currently
active stations broadcast, then the block continues until all these 
stations successfully dispatch their packets.
The protocol requires all the stations to listen to the channel 
all the time to be aware of the traffic and to be able to decide 
when a new block begins.

It also follows directly by the Kaplan criterion (see Section~\ref{tools}) 
that this version of Aloha with control is not stable for the arrival 
rate greater than~$1/e\approx 0.368$, notice that the downward transitions 
are uniformly bounded.
One can show that the protocol is actually transient.

Consider a more general scenario when the process of arrival of new 
packets has distribution $\langle \lambda_i\rangle$, that is, 
$\lambda_i$ is the probability that exactly $i$~new packets arrive in a step.
We consider such algorithms in which each among the~$n$ stations active at a
step enables itself with some probability~$f(n)$,
independently across steps and stations.
Then $a(n)=(1-f(n))^n$ is the probability of a silent channel, and
$b(n)=n\cdot f(n) \cdot (1-f(n))^{n-1}$ is the probability of a successful
broadcast.
Fayolle, Gelenbe and Labetoulle~\cite{FayolleGL77} showed that if the limit 
\[
d=\lim_{n\rightarrow\infty}\lambda_1\cdot a(n) +\lambda_0\cdot b(n)
\] 
exists then the following holds: for $\lambda<d$ the protocol is stable 
and for $\lambda>d$ it is unstable.
Consider such protocols for which both limits 
$\lim_{n\rightarrow\infty}f(n)$ and $\lim_{n\rightarrow\infty}n\cdot f(n)$
are finite, as well as~$d$, and also suppose that
$\sum_{i\rightarrow\infty}i\cdot \lambda_i=\lambda>\lambda_1$.
As a corollary, we obtain that for such a protocol 
to be stable it is necessary that both of the following conditions:
\begin{enumerate}
\item
$\lim_{n\rightarrow\infty} f(n) = 0$ \ ; 
\item
$\lim_{n\rightarrow\infty} n f(n) > 0$ \ ;
\end{enumerate}
hold. 
Namely, if the first condition does not hold then $d=0$.
On the other hand, if the second one does not hold then $d=\lambda_1<\lambda$. 
In both cases the protocol is unstable.

\Paragraph{Static model.}

Pippenger~\cite{Pippenger81} showed that the capacity of a channel with 
the size-detection mechanism equals one under the static scenario.  
The model considered in~\cite{Pippenger81} is that of finitely many messages 
with arrival times in the interval $(0,1)$ generated by a Poisson process, 
but the approach translates in a natural way into the static infinitely-many 
users model with Bernoulli arrivals, which we consider in this exposition.
The algorithm developed in~\cite{Pippenger81} is not constructive, 
the proof is by the probabilistic method~(\cite{AlonS08}).
The algorithm operates in stages, each broadcasts the messages of the stations
in a segment of $\Theta (N)$ stations.
Consider such a consecutive segment~$X$, and let~$N$ be the exact number 
of messages in the stations in~$X$, which can be found in one step by the 
size detection.
Let $B=2\lceil N (\log N)^{1/2}\rceil$.
The segment~$X$ is conceptually partitioned into $B$ subsegments 
$X_1,\dots,X_B$, of approximately equal sizes.
Next the algorithm proceeds through two phases: during the first one 
the number $N_i$ of messages in each segment $B_i$ is found, during the next 
one the actual broadcasting is performed.
The first phase is to gather information, although incidental successful
transmissions may happen, and the fact that this can be done in
time $\cO(N/(\log N)^{1/2})$ is the crux of the proof.
The second phase takes time $N+\cO(N/(\log N)^{1/2})$.
The design of phase one is based on the following combinatorial fact:
for a natural number $L$ there is a $K\times L$ binary matrix $F$, where 
$K=\cO(L/\log L)$ is a natural number, such that each sequence $G=\langle
G_1,\ldots,G_L\rangle$ with the property $G_1+\ldots+G_L\le L$
is uniquely determined by the vector $FG$ of length $K$.
The existence of such a matrix is proved by random coding
and extends a result of Erd\"os and  R\'enyi~\cite{ErdosR63}.
The first phase begins with determining the numbers of messages
$N^{(1)}=N_1+\dots+ N_{B/2}$ in $X^{(1)}=X_1\cup\ldots\cup X_{B/2}$ 
and $N^{(2)}=N_{B/2+1}+\dots+ N_{B}$  in 
$X^{(2)}=X_{B/2+1}\cup\dots\cup X_{B}$.
If some of them are at most one, what may happen with exponentially small
probability, then the respective $X^{(i)}$ are handled either directly or
recursively.
Otherwise the numbers $N_i$ are found without causing any successful 
broadcasts.
To this end the combinatorial fact mentioned above
is used with $L=B/2$, so that $K=\cO(N/(\log N)^{1/2})$.
In the second phase the algorithm first performs successful broadcasts
by enabling the stations in $X_i$ for $N_i=1$, 
and handles the remaining segments $X_i$ recursively.
It takes $N+\cO(N/(\log N)^{1/2})$ steps to broadcast $N$ messages, so an 
arbitrarily high transmission rate can be attained by taking $N$ sufficiently
large.

\subsubsection{Full sensing protocols}

Next we consider protocols which rely on the ternary feedback.
First we present the randomly adaptive algorithms, then the tree 
and splitting ones.

The first randomly adaptive algorithm was proposed by 
Hajek and van Loon~\cite{HajekL82}. 
Each station maintains parameter $f$ and each active station broadcasts a
packet with probability $f$.
This parameter $f$ is not changed after a success, and is multiplied by 
a constant after each step without a successful broadcast, that is why it may
be called a {\em multiplicative\/} protocol.
These constants are as follows:
$1.518$ when the channel is silent, and $0.559$ in the case of a collision.
This protocol was proved by Hajek~\cite{Hajek82}
to be stable for $\lambda<e^{-1}\approx 0.368$.

Kelly~\cite{Kelly85} proposed randomly adaptive {\em additive\/} protocols, 
in which at each step~$t$ each station maintains parameter~$A(t)$ 
interpreted as the estimated current number of active stations. 
An active station performs a broadcast at step~$t$ with probability~$1/A(t)$.
This parameter~$A(t)$ is updated after a step by adding a suitable constant
among some fixed three ones, depending on the respective feedback from 
the channel.
Rivest~\cite{Rivest87} proposed a specific additive
adaptive protocol in which parameter $A(t)$ is updated as follows:
if during step $t$ there is no collision then 
$A(t+1) = \max\{1,A(t)-1+\frac{1}{e} \}$ 
otherwise 
$A(t+1) = A(t)+\frac{1}{e-2}+\frac{1}{e}$.
Tsitsiklis~\cite{Tsitsiklis87} proved that this protocol is stable
for the arrival rate $\lambda<e^{-1}\approx 0.368$.

Another paradigm of full-sensing algorithms was proposed independently
by Capetanakis~\cite{Capetanakis79}, Hayes~\cite{Hayes78} and
Tsybakov and Mikhailov~\cite{TsybakovM78}. 
The algorithms they developed are usually called {\em tree algorithms}.
They operate as follows.
Let~$A$ be a positive integer interpreted as the arity of a tree.
The algorithm run by a station is either in a waiting phase or a 
conflict resolution phase, switching between them accordingly.
A {\em conflict resolution phase\/} starts after a conflict occurs.
A {\em waiting phase\/} starts after all the packets of the stations 
involved in a conflict resolution have been transmitted successfully. 
If a new packet is generated then it is assigned to a station,
which broadcasts it immediately if being in a waiting phase, otherwise it 
waits for the beginning of a waiting phase.
Conflict resolution is performed as follows. 
First each among the involved stations selects randomly and independently 
an integer in the segment $[1..A]$, we say that they {\em branch out}.
Then $A$ stages follow, the stations that selected number~$i$ participate 
in the $i$-th stage. 
They begin with broadcasting their packets. 
If a conflict occurs then it is resolved recursively, 
which may be interpreted as going down one level in a tree.
A stage terminates when all the participating stations have succeeded.
Parameter $A$ is often taken equal to~$2$. 
If the channel is quiet during all the branchings $1$~through $A-1$, then 
stage~$A$ need not be performed, it will certainly result in a collision,
hence the stations might branch out immediately.
Tree algorithms are stable for sufficiently small arrival rates,
namely their variants were shown to be stable for $\lambda <3/8$ 
in~\cite{TsybakovM78} and for $\lambda<0.430$ in~\cite{Capetanakis79}.
Notice that tree algorithms are not FCFS.

A modification of the tree-algorithm paradigm, called 
{\em splitting algorithm}, was proposed independently by Gallager 
and by Tsybakov and Mikhailov (see~\cite{Gallager85}).
This algorithm is FCFS.
The idea is to abandon the underlying tree structure and refer 
to the stations directly by their activation times.
The stations join the conflict resolution in the order of activation.
This  may be interpreted as if  they were stored in a stack.
Suppose that at step~$k$ the protocol has just successfully transmitted 
all the messages that arrived before some~time $T(k)$.
All the messages that arrived in the time period~$[T(k),T(k)+a(k)]$ are 
broadcast in slot~$k$.
If there is a collision then this time interval is partitioned into
subintervals, namely into {\em the left interval\/} $[T(k),T(k)+b(k)]$ 
and {\em the right interval\/} $[T(k)+b(k), T(k)+a(k)]$, where~$0<b(k)<a(k)$.
The stations from the left interval broadcast  first.
If a collision happens again then the left interval is subdivided into two
subintervals.
It is now when the departure from the tree-algorithm paradigm happens.
Namely, only these two subintervals obtained from the left interval 
are remembered, the stations from the right interval are put back on 
the stack, and are not distinguished from the other stations still there.
Otherwise, if there is no collision in the left interval, then the right 
interval is considered in a similar fashion.
The function~$a(k)$ is of the  form~$\min [c,k-T(k)]$, where the constant~$c$
is taken to be~$c=2.6$ to maximize the stable transmission rate.
The simplest function~$b(k)$ to be considered is the one that halves 
the interval~$[T(k),T(k)+a(k)]$, this determines the {\em binary\/} 
version of the splitting algorithm.
Binary splitting algorithm was shown to be strongly stable for 
$\lambda<0.4871$.

This binary algorithm was later streamlined by Mosely and Humblet, 
and Tsybakov and Mikhailov, who showed that by partitioning the interval 
in an optimal way increases the tenacity to~$0.4878$ (see~\cite{MoselyH85,TsybakovM80,Verdu86}).
Vvedenskaya and Pinsker~\cite{VvedenskayaP83} developed a protocol with tenacity 
still better by $3.6\times 10^{-7}$.
More on the analysis of parameters of the tree and splitting algorithms can be
found in~\cite{FayolleFHJ85,GreenbergFL87,MathysF85}.
In particular, they are strongly stable for sufficiently small 
arrival rates.

One might wonder if the best full-sensing protocols are FCFS.
This hypothesis was shown to be doubtful by Vvedenskaya and Pinsker~\cite{VvedenskayaP83}
whose protocol is not FCFS.
The issue of optimality among subclasses of full-sensing algorithms was also
studied by Molle~\cite{Molle81-techreport} and then by Panwar, Towsley and Wolf~\cite{PanwarTW85}.
In particular, they considered {\em nested FCFS\/} algorithms which enable
stations in the following conservative way:
\begin{enumerate}
\item
If no set of stations is known in a step to contain more than one
message and the set of stations~$C$ enabled in the next step contains 
station~$s$ then set~$C$ contains also all the stations preceding~$s$ 
whose status, with respect to holding a message, has not been 
clarified yet (this is the FCFS property). 
\item
If a set~$B$ of stations is known in a step to contain more than one 
message then in the next step a subset of $B$ is enabled 
(this is the property of being nested).
\end{enumerate}

A method to construct an optimal nested FCFS algorithm for a given probability
to hold a message (Bernoulli arrivals for static model) has been 
developed in~\cite{PanwarTW85}. 
It is based on the Markovian decision theory, in particular
on the value iteration algorithm (see~\cite{Howard62-book}) and the Odoni bound 
(see~\cite{Odoni69}).
Panwar, Towsley and Wolf~\cite{PanwarTW85} developed a scheme of {\em mixing
algorithms}, which are not FCFS, and showed that they have higher 
transmission rates than optimal nested FCFS ones, 
for certain small probabilities to hold messages.
To obtain such an algorithm, run an optimal nested FCFS algorithm 
augmented with the following trick. 
Suppose that we have identified a set $\{ s_1,s_2,s_3\}$ of three stations, 
of which we know only that at least two of them hold packets. 
Then instead of enabling either $\{ s_1\}$ or $\{ s_1,s_2\}$, as a 
FCFS algorithm would do, enable $\{s_1,s_4 \}$, where station~$s_4$ 
is the next station after $s_3$, and which may or may not hold a message.
This approach works because, for sufficiently small probability to hold 
a message, station~$s_4$ will likely turn out not to hold a message, 
so with a single try we will eliminate two stations.

\subsubsection{Acknowledgement-based protocols}
\label{abp}

Now we consider protocols which rely on the weakest feedback, 
that of a network without collision detection.
They can be run on any single-hop network, they simply ignore
the information from the channel, if any,  obtained while staying idle.
Such protocols usually erase their history after a successful broadcast, so
they are acknowledgement-based. 
Because of that, we will not distinguish between possibly more general
arbitrary protocols for the channel without collision detection and
acknowledgement-based protocols. 

The oldest protocols, namely Aloha and the binary exponential backoff,
are in this class.
The infinitely-many users model was initially considered with a hope
to show in a clean setting that these two protocols
were stable, at least for sufficiently small arrival rates.
First, it was realized that the expected packet delay in Aloha is 
infinite (see~\cite{Abramson77,Kleinrock-book75-76}).
Then it was shown that Aloha is unstable for any arrival rate 
(see~\cite{FayolleGL77}). 
This follows directly from the Kaplan's criterion (see Section~\ref{tools}). 
Namely, first notice that the downward transitions are uniformly bounded.  
Secondly, let $0<f<1$ be the probability that a station with a backlogged 
packet performs a broadcast in a step, and let~$i$ be the number of 
such stations.
Then the drift is at least
\[
2\cdot\bigl (1-e^{-\lambda}-\lambda e^{-\lambda}\bigr)
-\bigl( e^{-\lambda}\cdot i\cdot f\cdot (1-f)^{i-1}\bigr) .
\]
So it is positive and bounded away from zero by a constant for sufficiently
large~$i$.

Rosenkrantz and Towsley~\cite{RosenkrantzT83} proved that Aloha is 
transient, by their criterion based on the martingale convergence theorem 
(see Section~\ref{tools}).
This result was later strengthened by Kelly~\cite{Kelly85} who showed that
with probability one the channel transmits a finite number
of messages and then becomes jammed forever; moreover one can
show that the expected time for this to happen is finite.
Hence Aloha is not weakly stable for any arrival rate.
The reasoning in~\cite{Kelly85} was as follows.
Let~$N(t)$ be the number of stations with backlogged packets at step~$t$,
and~$Z(t)$ be the number of broadcasts at step~$t$.
Let~$p(n)$ be the probability that the event~$Z(t)\le 1$ happens 
before the backlog increases from~$n$.
It satisfies the equation 
\[
\Pr (Z(t)\le 1 \mid N(t)=n) = p(n)\cdot (1-\Pr (N(t+1)=n, Z(t)>1\mid N(t)=n))\ .
\]
This implies, by straightforward calculations, that 
\[
p(n) \sim \frac{nf(1-f)^{n-1}}{1-e^{-\lambda}}\ ,
\]
with $n\rightarrow\infty$.
Let~$r(i)$, for~$i=1,2,3\ldots$, be the times at which the backlog reaches
record values, that is, $r(1)=1$, and 
\[
r(i+1)=\min \{ t>r(i) : N(t)>N(r(i))\}\ .
\]
Then $\sum_{i\ge 1} p(N(r(i)))<\infty$ and by the Borel-Cantelli lemma 
we obtain that with probability one the channel will jam itself 
forever after a finite number of steps.

Kelly and MacPhee~\cite{Kelly85,KellyM87} generalized this approach to cover 
acknowledgement based protocols, a sketch of their analysis follows.
Let us fix an acknowledgement protocol.
For the purpose of argument let us assume that the channel is externally 
jammed from the very beginning so that no packet is successfully broadcast.
Let~$g(k)$ be the probability that a station broadcasts a packet in the 
$k$-th step since it was generated, conditioned on the event that 
the previous attempts to broadcast this packet failed.
Let  $s(t)=\sum_{k=1}^{t}g(k)$.
The number of transmissions made in step~$t$ is Poisson with parameter
$\lambda\cdot s(t)$.
Hence the probability that less than two broadcasts are made in a slot is 
\[
u(t)=(1+\lambda s(t))\exp(-\lambda s(t))\ .
\]
The expected number of such slots is
\[
U(\lambda)=\sum_{t\ge 1} u(t)\ .
\]
The assumption about the external jamming can be removed by considering
arbitrarily small additional Poisson traffic.
Function $U(\lambda)$ is nonincreasing and it may be equal to infinity for
sufficiently small~$\lambda$.
It categorizes the arrival rates as follows: if $U(\lambda)<\infty$ then with
probability one the channel has only finitely many successful transmissions,
and if $U(\lambda)=\infty$ then the protocol is weakly stable.

Define the critical arrival rate as follows: 
\[
\lambda_c=\inf \{ \lambda >0: U(\lambda)<\infty\}\ .
\]
Hence if $\lambda>\lambda_c$ then the number of packets successfully 
transmitted is finite with probability one, and if $\lambda<\lambda_c$ then the
protocol is weakly stable.
In particular, if $s(t)=o(\ln t)$ then $\lambda_c=0$,
and if $\ln t=o(s(t))$ then $\lambda_c=\infty$.
Consider specific protocols as examples.
For instance, we have $s(t)\sim f\cdot t$ for the Aloha scheme, 
hence $\lambda_c=0$.
The estimation is $s(t)\sim \log_a t$ for the exponential backoff with 
factor~$a$,  so we have~$\lambda_c=\ln a$; in particular
$a=2$ and $\lambda_c\approx 0.693$ for the binary exponential backoff.
Finally, any backoff protocol, with the backoff rate asymptotically smaller 
than that of an exponential backoff, like a polynomial one, 
is not weakly stable.

Aldous~\cite{Aldous-TIT87} showed that the binary exponential backoff is
not stable for any arrival rate.
More precisely, he showed that this protocol is transient and has zero
throughput.
Combining the results from~\cite{Aldous-TIT87,Kelly85,KellyM87} we can see that it is possible 
for a backoff protocol to be weakly stable, for sufficiently small traffic,  
and simultaneously non-recurrent and of zero throughput.
This also shows that for the infinitely-many users model the exponential 
backoff protocols may be considered superior to the polynomial backoffs, 
in the sense that the former can be weakly stable if the traffic is 
sufficiently small, while the latter never are.
This is in contrast with the finitely-many users model, see 
Subsection~\ref{fum-dc}, where arguments for the opposite preferences 
are given.

We present a sketch of the approach  from~\cite{Aldous-TIT87}.
The interpretation used is in terms of pebbles and boxes, with messages 
called pebbles.
At each step new messages (pebbles) are placed in box number zero,
and a pebble in box~$i$ is either moved to box $i+1$ 
with probability~$2^{-i}$ or remains in~$i$, 
independently for different pebbles.
This can be interpreted as a Markov chain, with the numbers of pebbles in the
boxes giving the state.
It has the stationary distribution which places a number of pebbles in box~$i$
with the Poisson distribution with parameter $\lambda 2^i$.
Start the chain with this distribution.
Let $Y_i(t)$ be the number of pebbles in box~$i$.
This process $Y(t)=\langle Y_1(t), Y_2(t),\ldots\rangle$ models 
an externally jammed channel.
There is another process~$X(t)$, which uses a coloring scheme.
A new pebble is red, and if exactly one red pebble is moved in a step 
then it is recolored white.
The number of white pebbles in box~$i$ at step~$t$ is denoted by $X_i(t)$,
this defines the process $X(t)=\langle X_1(t), X_2(t),\ldots\rangle$, which 
captures the real behavior of the channel, because the white pebbles 
correspond to successful transmissions. 
For the externally jammed channel the expected number of packets with $i$
unsuccessful transmissions is $\mE \, Y_i=\lambda 2^i$.
The real channel is represented by $X(t)$. 
One of its properties is that if $X_i(t)\ge \lambda 2^i$ then 
a positive chance exists than the backlog of packets increases. 
To argue more precisely, let us introduce the notation
\[
f(x) = \sum_{i\ge 1} x_i\cdot 2^{-i}
\]
for a sequence of nonnegative integers 
$x=\langle x_0, x_1, x_2,\ldots\rangle$.
A key technical observation is that the probability of the event that some
pebble is recolored in the step~$t+1$, conditioned on the state being 
represented by~$x$, is at most $2\exp(-f(x)/2)$, which is proved by the
Chernoff bound (see~\cite{McDiarmid98,MotwaniR95}).
Then it is shown that~$f(X(t))$ converges to infinity with the probability one,
as~$t$ goes to infinity, what yields the result by a probability estimate.

Kelly and MacPhee~\cite{KellyM87} asked the question if there exists 
an acknowledgement based protocol which is recurrent 
in the infinitely-many users model with Poisson arrivals.
This was answered in the affirmative by Goldberg, MacKenzie, Paterson, 
and Srinivasan~\cite{GoldbergMPS-JACM00}, who developed a protocol with a constant  
expected delay, and hence strongly stable, for arrival rates smaller 
than $1/e$.
We give an overview of the protocol.
There is a conceptual tree of infinitely many levels, and infinitely many nodes
at each level.
With each node $v$ there is associated a {\em trial set}, denoted by
$Trial(v)$, of time slots.
The first and last elements of $Trial(v)$ are denoted as $L(v)$
and $R(v)$, respectively.
The number of elements in $Trial(v)$ is referred to as the 
{\em size\/} of~$v$.
The trial sets make a partition of the positive integers (are disjoint and
cover all these integers) with these properties:
\begin{enumerate} 
\item
If either $u$ is a proper descendant of $v$, or $u$ and $v$ are on the same
level with $u$ to the left of~$v$, then $R(u)<L(v)$.
\item
The sizes of nodes at the same level~$i$ are all equal to $br^i$,
for some constant parameters $b$ and~$r$.
\end{enumerate} 
The arity of the tree is some positive integer~$k$, which is a parameter
satisfying $k>r$.
The protocol operates as follows.
When a message~$m$ arrives at time step $t$ then it is assigned to the 
leaf $v$ which is the leftmost one among those with the property $L(v)>t$.
As soon as $m$ is assigned to a node $v$ then it picks a time $t$ uniformly at
random from $Trial(v)$, and is attempted to be sent onto the channel at time
step~$t$.
If a failure occurs then message~$m$ is moved immediately to the parent of~$v$. 
This defines the way message~$m$ keeps moving up the tree along the ancestor 
links until eventually it is broadcast successfully.

Notice that the protocol described above resembles the unmodified exponential
backoff protocols in that it uses a partition of time steps into disjoint 
subsets $S_{i,t}$ of sizes growing exponentially in $i$. 
Messages that arrive at time step~$t$ jump from one subset $S_{i,t}$ to the
next one $S_{i+1,t}$ after having failed to use the channel at 
a random time step in $S_{i,t}$.
The efficiency of the protocol stems from the fact that groups
of many messages arriving together are spread among a number of leaves and
hence handled independently; this is because the trial sets of leaves
are not contiguous segments of time steps but are intertwined with the trial
sets of the nodes from higher levels.

\section{Finitely-many users models}

In this section we continue our discussion of the single-hop networks,
concentrating on the case of finitely many stations, which may be 
considered as more realistic.
Also it allows to study natural problems for the static arrivals which have no
counterparts in the infinitely-many users model, namely when the packets are
distributed arbitrarily among the stations.
These problems are considered in the next subsection.
We assume the ternary feedback, unless stated otherwise.

\subsection{Static case}

Let $n$ be the total number of stations, and suppose some $k$ of them have
messages.
The static selection problem is to broadcast just {\em any\/} single message
successfully.
The all-broadcast problem is to make {\em all\/} the~$k$ stations with 
messages to broadcast successfully.
Notice first that if the value of~$k$ is known to the participating stations
then they may run controlled Aloha, that is, each would broadcast with
probability $1/k$ in a step, independently, and a successful broadcast will
happen after the expected time $\cO(1)$.
Hence we assume in what follows that the number~$k$ is not known to the
stations.

Willard~\cite{Willard-SICOMP86} developed a randomized protocol for the  
static selection problem. 
It is in two versions.
The first one covers the scenario when number~$n$ is known,
it has the expected time $\lg \lg n + \cO(1)$.
The second version assumes that $n$ is unknown and terminates in the expected
time $\lg\lg k +\cO(\lg\lg\lg k)$.

We present a brief overview of the approach in~\cite{Willard-SICOMP86}.
Suppose first that number~$n$ is known.
The algorithm is in two phases.
The underlying idea of the first phase is to perform a binary search on 
a space of size~$\log n$ to estimate the number~$k$ of stations with messages. 
This is done by adaptively adjusting the individual probabilities to 
broadcast.
More precisely, let~$K$ be an integer, and let TEST($K$) be a procedure 
in which each stations performs a broadcast in a step with the 
probability~$1/K$; it returns either {\tt silence} or {\tt success} or 
{\tt collision}.
Let $\lg ()$ be the binary logarithm.
The binary search of the first phase is as follows:
\begin{enumerate}
\item
Set $L=0$ and $U=\lceil \lg n\rceil+1$;
\item
{\tt while} $L\neq U$ {\tt do}:
\begin{enumerate}
\item
set $i=\lceil (L+U)/2\rceil$ and $K=2^i$;
\item
set $t=$TEST$(K)$;
\item
if $t= \mbox{\tt success}$ then terminate the whole protocol;
\item
if $t=\mbox{\tt silence}$ then $U=i$;
\item
if $t=\mbox{\tt collision}$ then $L=i$.
\end{enumerate}
\end{enumerate}
Then the second phase is performed, in which we plug the latest~$K$
as the estimate of the number of stations into controlled Aloha 
and hope for the best.
The first phase is never performed for more than $\lceil \lg\lg n\rceil$
steps.
A crucial thing in the analysis is to show that the expected 
number of steps during the second phase is~$\cO(1)$.

Consider now the scenario when the number of stations~$n$ is unknown.
The binary search cannot be applied, but it can be replaced by the 
search algorithm of Bentley-Yao~\cite{BentleyY76} yielding a second version of
the algorithm, which has the expected time $\lg\lg k +\cO(\lg\lg\lg k)$. 

Willard~\cite{Willard-SICOMP86} also proved a matching lower bound to show the
optimality of the first version of the protocol for a known number~$n$.
However the  considered protocols were assumed to be {\em fair\,}: 
all the stations that toss coins to decide if to broadcast use 
coins with the same probability of heads to come up.
The following was shown: there is an absolute constant~$c$ such that any fair
protocol has the expected time at least $\lfloor \lg\lg n \rfloor -c$,
for some~$k$.

The algorithm of Willard operates on a single-hop radio network with collision
detection.
Kushilevitz and Mansour~\cite{KushilevitzM-SICOMP98} showed a lower bound $\Omega(\log n)$
for the static selection problem if collision detection is not available.
This yields an exponential gap between the two models.

Next we consider a related problem of finding maximum:
suppose some~$k$ among~$n$ stations hold keys, we need to find the maximum 
among them with respect to some ordering among keys.
A combination of the binary search with respect to the ordering of keys
and the selection algorithm of Willard gives an algorithm with
the expected time $\cO(\log\log k\cdot\log k)$.
Martel and Vayda~\cite{Martel-IPL94,MartelV88} showed that this can be streamlined to 
$\cO(\log k)$.
To this end a binary search is applied, which proceeds by creating 
a sequence of subsets $\langle A_i \rangle$ of diminishing sizes.
The next subset $A_{i+1}$ of a current subset $A_i$ is created by 
partitioning $A_i$ according to the key of a random element in~$A_i$.
Such an element is selected by running a modified selection procedure.
The idea is to start the first phase from the estimate of the size of
the previous subset.
This results in an amortization of the binary search among the consecutive
successful broadcasts.

\subsubsection{Deterministic solutions}

First we consider the selection problem.
Notice that the tree algorithm, adapted to a finite number 
of stations with identification numbers, solves the selection problem 
among $n$ stations in $\lceil \lg n\rceil$ steps.
Martel and Vayda~\cite{MartelV88} showed a lower bound of at least $\lg (n-k)-c$
steps on any deterministic algorithm, for a certain constant~$c$.
Both this lower bound and the randomized protocols of Willard 
demonstrate that randomization allows to develop selection protocols 
with asymptotically better expected performance than the worst-case 
performance of any deterministic protocol.

Next we consider the all-broadcast problem.
For unknown~$k$ and known~$n$, the problem can be also solved by 
the tree algorithm, with the worst-case time bound $\cO(k+k\log(n/k))$.
Koml\'{o}s and Greenberg~\cite{KomlosG-TIT85} developed a nonadaptive deterministic 
algorithm to solve the all-broadcast problem in time $\cO(k+k\log(n/k))$, 
where both numbers~$n$ and~$k$ are known.
The proof is nonconstructive, by the probabilistic method.
Greenberg and Winograd~\cite{GreenbergW-JACM85} proved a lower bound of $k+\lg(n/k)$ steps
on the time to solve the all-broadcast problem deterministically, which they
generalized to a lower bound $\Omega(k(\log n)/(\log k))$.

\subsubsection{Optical communication}

A channel even weaker than a single-hop radio network without collision 
detection can be considered,  we call it {\em optical channel}.
It operates as follows.

\begin{myquote}
{\sf In an optical channel:}

A station that performs a broadcast onto the channel, and is the only 
station broadcasting in a step, receives the message 
{\tt success} as feedback.

If at least two stations broadcast simultaneously in a step then each of 
them can hear only the background noise.

A station that does not attempt to broadcast in a step can hear only the
background noise. 
\end{myquote}

The property that none of the idle stations receive the messages sent
successfully to the channel by other stations is what distinguishes 
the optical channel from the single-hop radio network without 
collision detection.
Notice that the stations that actually perform a broadcast 
in a step can recognize a collision when it happens.

A problem similar to the all-broadcast problem for the multiple access channel, 
when each of some~$k$ among the~$n$ stations needs to broadcast its message 
on the optical channel, and each of the stations with messages does know 
the number~$k$, has been called the 
{\em control tower problem\/} in~\cite{MacKenziePR98}.
Ger\'eb-Graus and Tsantilas~\cite{Gereb-GrausT92} developed a randomized protocol 
solving this problem in time $\cO(k+\lg n\lg k)$ with the  probability 
polynomially close to one.
MacKenzie, Plaxton and Rajaraman~\cite{MacKenziePR98} proved the following lower bound: 
if a randomized algorithm solving the control tower problem operates in time
$T(n,k)$ with probability at least $1-n^{-3/4}$ then 
$T(n,k)=\Omega(\log k\log n)$.
The control tower problem can be solved in the expected time~$\cO(k)$.

The OCPC model of optical communication can be interpreted as consisting
of some~$n$ stations, each with a dedicated optical channel, 
the owner of a channel receives messages broadcast on it successfully.
A basic communication problem for the OCPC model is to 
{\em realize an $h$-relation}: each node is a source of at most~$h$ packets, 
and also a destination of at most~$h$ packets, we need to deliver the packets
successfully.
Goldberg, Jerrum, Leighton and Rao~\cite{GoldbergJLR97} developed a randomized routing
protocol realizing $h$-relations in the expected time $\cO(h+\log\log n)$.
A routing algorithm is {\em direct\/} if nodes may send packets 
only to their destination nodes.
The algorithm presented in~\cite{GoldbergJLR97} is not direct.
A direct randomized protocol to realize~$h$ relations was developed 
in~\cite{Gereb-GrausT92}, it operates in time $\Theta(h+\lg n\lg h)$ with high 
probability; actually  only $h\ge \log n$ was considered and the obtained time 
was $\Theta(h+\log n\log\log n)$.
A lower bound on direct algorithms was shown in~\cite{GoldbergJLR97}: 
a direct randomized algorithm that can realize any $2$-relation with the
success probability of at least $1/2$ needs time $\Omega(\log n)$ 
on some $2$-relation.
An $\Omega(h+\sqrt{\log\log n})$ lower bound, for routing $h$-relations 
in the OCPC model, was shown by Goldberg, Jerrum and MacKenzie~\cite{GoldbergJM98}, 
with no restriction on algorithms, in particular it covers indirect 
algorithms.

\subsection{Dynamic arrivals}
\label{fum-dc}

Let the number of stations be denoted by $n$.
The average arrival rate at station~$i$ is $\lambda_i$,
and $\lambda=\sum_{1\le i\le n}\lambda_i$ is the total arrival rate.

\subsubsection{Aloha}

If arrivals are diversified among the stations, then we need 
the probability of performing a broadcast in Aloha also diversified,
that is, station~$i$ performs a broadcast in a step with some 
probability~$f_i$, provided its queue is nonempty, independently of 
the other steps and stations.
Tsybakov and Mikhailov~\cite{TsybakovM79} considered arbitrary arrival processes,
not necessary Bernoulli, and showed that Aloha was stable for certain 
configurations of parameters. 
For more on the topic see~\cite{SaadawiE81,Szpankowski88,SzpankowskiR88}.
These results are in contrast with the situation for the infinitely-many 
users model where Aloha is not even weakly stable
(see Subsection~\ref{abp}).
However this is not that surprising if we realize that with the number of
station~$n$ fixed and known to all the stations, using $f_i=\Theta(1/n)$
yields a process that is quite similar to controlled Aloha, 
hence it might be expected to be stable, at least for certain arrival rates.

Consider the Markov chain~${\cal M}_o$ which has the sequences of 
sizes of the buffers $\langle q_1,\ldots,q_n\rangle$ as its states.
This chain is not state homogeneous, in the sense
that the probability of moving from state~$a$ to~$b$ by vector~$c=b-a$ 
does not depend only on~$c$.
The approach in~\cite{TsybakovM79} was by way of considering another 
state homogeneous Markov chain, which was suggested by the observation 
that Aloha has its worst time when all the buffers are nonempty.
First let us present intuitions of this approach, inspired by~\cite{TsybakovM79}.
For the simplicity of calculations, let us consider the special case when the
arrivals are Bernoulli and distributed evenly among the stations, that is 
$\lambda_i=\lambda/n$, and also where the probabilities $f_i=f/n$ are all
equal, for $i<n$.
Define the protocol {\em Aloha with jamming\/} to run like
Aloha but if a station has an empty buffer then let it immediately
generate itself a dummy message and run Aloha until the packet with this
message is successfully broadcast. 
It is intuitively clear that if the Aloha with jamming is stable so is 
the original Aloha.
We compute the drift in order to apply the Pakes' criterion.
It is equal to the average arrival rate minus the average departure rate.
The departure rate is
\[
n\cdot \frac{f}{n}\cdot \Bigl(1-\frac{f}{n}\Bigr)^{n-1} e^{-\lambda}
\approx
f e^{-f}e^{-\lambda}\ .
\]
However the arrival rate is not just $\lambda$ but also the average arrival 
rate of the dummy packets. 
This additional rate can be estimated directly  by finding the stationary 
distribution of the Markov chain, which requires solving a linear recurrence 
with constant coefficients. 
After combining all this we obtain a bound on the arrival rate which guarantees
ergodicity,  details are omitted.

The approach presented in~\cite{TsybakovM79} was direct, by way of generating 
functions and systems of equations to determine probabilities, 
rather than by applying any ergodicity criteria.
The authors considered Markov chains~${\cal M}_k$, for $k=1,\dots,n$, 
with the state space $0,1,2,\ldots$, where state~$i$ corresponds to 
the total number of packets in the buffer of the $k$-th station, 
and the probability~$p_{ij}$ of going from~$i$ to~$j$ is defined 
as the probability of changing the buffer of the $k$-th station by~$j-i$ 
packets in chain~${\cal M}_o$, conditioned on all the buffers not being empty. 
It was shown in~\cite{TsybakovM79} that chain~${\cal M}_k$ is ergodic if the 
inequality~$\lambda_k<\gamma_k$ holds, where 
\[
\gamma_k=f_k\prod_{i\ne k}(1-f_i)\ .
\]
Then it was proved that if all the inequalities $\lambda_k<\gamma_k$ hold, 
for $1\le k\le n$, then Aloha is stable.
As a corollary, it follows that if $\lambda<e^{-1}$ then Aloha is stable,
for probabilities $f_i=e\lambda_i(1+e)^{-1}$.

\subsubsection{Backoff protocols}

Recall that neither polynomial nor exponential backoff protocols are 
stable in the infinitely-many users model (see Subsection~\ref{abp}).

H\aa stad,  Leighton and  Rogoff~\cite{HastadLR-SICOMP96} showed that the
binary exponential backoff is not stable 
if the arrival rates at stations are equal and their sum $\lambda$ 
exceeds $0.567+(1/(4n-2))$.
The method used was by considering the following test function: 
\[
F(\langle q_1,\dots,q_n,b_1,\ldots,b_n\rangle)
=
(2n-1)\sum_{i=1}^n q_i +\sum_{i=1}^n 2^{b_i} -n ,
\]
where $\langle b_i\rangle $ are backoff counters, and $\langle q_i\rangle$
are queue sizes.
It was shown in~\cite{HastadLR-SICOMP96} that this function is expected to grow  
by at least a fixed positive amount, from which the instability follows.
This result was strengthened asymptotically in~\cite{HastadLR-SICOMP96} to show that if the
arrival rate is bounded away from~$1/2$ by any constant~$c>1/2$
then the system is not stable for the number of stations sufficiently large,
depending on the constant~$c$.

It was also shown in~\cite{HastadLR-SICOMP96} that any linear or sublinear polynomial
backoff protocol is unstable for any arrival rate and sufficiently 
large number of stations.
The proof is by investigating the behavior of the following test function
\[
F(\langle q_1,\dots,q_n,b_1,\ldots,b_n\rangle)
=
n^{3/2} \sum_{i=1}^n q_i - \sum_{i=1}^n (b_i+1)^{a+1}\ ,
\]
where $h_i=i^{-a}$ are the backoff probabilities.

On the other hand, Goodman, Greenberg,  Madras and  March~\cite{GoodmanGMM88} 
showed that, for Poisson arrivals, the binary exponential backoff 
is stable when $\lambda<n^{-a\log n}$, for a certain constant $a>0$.
They also studied the special case of two stations in detail, with a 
specific bound on~$\lambda$.
Al-Ammal, Goldberg and MacKenzie~\cite{Al-AmmalGM-TCSy01} improved this result by showing
that the binary exponential backoff is stable for~$\lambda$ 
at most~$\cO(n^{-\delta})$, for~$\delta>0.75$. 
This was done by applying the generalized Foster's criterion
which employs two functions (see~Section~\ref{tools}).
Unfortunately the lower bound on $\lambda$ decreases to zero with 
$n\rightarrow\infty$.

Surprisingly enough, H\aa stad,  Leighton and  Rogoff~\cite{HastadLR-SICOMP96} showed that 
any superlinear polynomial backoff algorithm  is strongly stable for any
arrival rate smaller than one.
The proof was by investigating the properties of the test function
\[
F(\langle q_1,\dots,q_n,b_1,\ldots,b_n\rangle)
=
\sum_{i=1}^n q_i + \sum_{i=1}^n (b_i + 1)^{a+\frac{1}{2}}-n\ ,
\]
where the backoff rate has the form $(x+1)^{-a}$, for $a>1$, as a function
of~$x$.
Then sums of the form~(\ref{eq1}) (given in Section~\ref{tools}) 
of the values of $F$ over some states of the chain were investigated, 
and the respective criterion was applied.

This shows that exponential backoff protocols are not necessarily 
better than polynomial ones, but rather vice versa, what
is in contrast with the infinitely-many users model, see Subsection~\ref{abp}.
An essential property of backoff protocols in the finitely-many users 
model is that if a station broadcasts successfully then it keeps sending 
messages until eventually a collision occurs, so a success allows 
the station to grab the channel for a possibly substantial amount of 
time and hence to load off a lot of messages from its queue. 
This effect needs to be tuned to the rate of backing off,
that is why linear backoff is too slow, quadratic just perfect 
and exponential rather too fast.
The effect of holding the channel over a significant period of time cannot
happen in the infinitely-many users model because a station dies after 
a success.

Goldberg and MacKenzie~\cite{GoldbergM99} considered a system of clients 
and servers, each server as a multiple access channel. 
Clients generate requests to servers with some Bernoulli distributions.
The client-server request rate is the maximum, over all the pairs of a 
client and a server, of all the request rates associated with either 
the client or the server.
It was shown in~\cite{GoldbergM99} that any superlinear polynomial backoff 
protocol is strongly stable if the request rate is smaller than one.
This result can be applied to routing in optical networks, like the OCPC.

\subsubsection{Short-delay protocols}

The proof of a strong stability of polynomial backoff protocols 
given in~\cite{HastadLR-SICOMP96} yields exponential upper bounds on the waiting time in 
the number of stations.
It was also shown in~\cite{HastadLR-SICOMP96} that for any superlinear polynomial backoff 
protocol the expected waiting time is superlinear, more precisely, 
if the backoff rate is $f(x)=(1+x)^{-a}$ then the lower bound 
is~$\Omega(n^{\frac{a+1}{a}})$.

Raghavan and Upfal~\cite{RaghavanU-SICOMP98} developed an acknowledgement-based  
protocol which is strongly stable for sufficiently small arrival rates and
which has $\cO(\log n)$ expected delay.
This was the first strongly stable protocol with a provably
sublinear waiting time of messages.
An acknowledgement-based protocol, with a constant expected delay for arrival 
rates smaller than~$1/e$, was developed in~\cite{GoldbergMPS-JACM00},
it is an adaptation of the infinitely-many users 
version as discussed in Subsection~\ref{abp}.
Moreover, the stations need not be fully synchronized, provided that
each station survives for polynomially many steps every time it restarts.

The following fairly general lower bound was showed in~\cite{RaghavanU-SICOMP98}: 
for each acknowledgement based protocol, there is a number $0<a<1$ 
such that if the arrival rate $\lambda$ satisfies
$\lambda>a$ then the expected delay is $\Omega(n)$.
A sketch of the proof is as follows.
Suppose that the arrivals at stations are all equal 
to $\lambda/n$.
Let $X_t$ be the current state of the system at time~$t$, which is the 
total number of backlogged messages.
Suppose also that there is a stationary distribution, otherwise the system is
not positive recurrent and the bound holds.
In the remaining part of the proof the probabilities are calculated with 
respect to this distribution.
Let us consider the case $\Pr (X_t<n/2)>1/2$, otherwise the bound clearly holds.
Suppose that if $X_t<n/2$ then the probability
of a collision at step $t+n/3$ is at least~$q$.
The probability of a collision in a given step is at least $q/2$,
and the system cannot be stable if the arrival rate~$\lambda$ is greater
than~$1-q/2$.
So it is sufficient to show that the probability~$q$ is at least a constant.
There are at least $n/2$ senders with no packets with probability at
least~$1/2$, we call them {\em empty}.
The protocol determines the probability distribution $\langle p_i\rangle$ 
such that when a message arrives to an empty sender then it transmits 
this message at step $t+i$ with probability~$p_i$.
The number $\sum_{i=0}^{n/3} p_i$ is at least a constant $\beta$,
since otherwise the delay is at least $n(1-\beta)$.  
Thus the probability that any empty sender transmits at time $t+n/3$
is at least $\beta\lambda/n$, and the probability~$q$ of a collision 
is at least a constant.

\section{Multi-hop networks}

In the case of general multi-hop radio networks, various 
forms of dissemination of information can be studied.
The simplest one is that of {\em broadcasting}.
In this context it does not mean an attempt to send a single message
to the recipients of a station, as in the multiple access channel case, 
but rather the task of delivering a message originally stored by some
{\em source node\/} to all the other nodes in the network.
The source does not need an acknowledgement of the completion 
of broadcasting, otherwise the problem becomes {\em broadcasting with
acknowledgement}.
More complex communication problems concern many concurrent
instances of either point-to-point or broadcast communication tasks. 
We refer to all of them as {\em multiple communication}.
In particular, the problem when all the nodes need to perform broadcasting 
simultaneously is called {\em gossiping}:
initially each node knows its message, and all these messages
need to be acquired by every node.

We always assume that individual messages are of such size that each of them 
can be sent between adjacent nodes in one step.
In the context of multiple communication, 
a node may find it useful to transfer many messages simultaneously. 
It depends on the capacity of links if this is feasible.
If any number of individual source messages can be lumped together 
into a bigger one that still can be transmitted in one step 
then we say that {\em messages can be combined}.
If each of such messages has to be sent separately then we refer to the model
as that of {\em separate messages}.
When gossiping is considered, then the model assumed is that of combined
messages, unless stated otherwise.

All distributed protocols discussed in this section are designed for ad-hoc
networks, with nodes not assumed to know their adjacent ones.
The model is without collision detection, and the size of the network 
is known by the nodes, all this holds, unless stated otherwise.

Randomized algorithms are usually categorized as either {\em Monte Carlo\/}
or {\em Las Vegas} (see~\cite{MotwaniR95}).
The former perform their required tasks successfully with some positive
probability only, the latter terminate after having
completed their task. 
If an execution of a randomized algorithm results in
performing the required task then we say that it is {\em correct}.

We use the following notation in this section:
$n$ is the number of nodes, $D$ is the diameter of the network, 
$\Delta_{{\rm in}}$ is the maximum in-degree of a node, 
$\Delta_{{\rm out}}$ is the maximum out-degree of a node,
and $\Delta=\max\{\Delta_{{\rm in}}, \Delta_{{\rm out}}\}$
is the maximum degree.

\subsection{Communication algorithms}

We discuss three topics: randomized broadcasting, deterministic broadcasting, and multiple communication.

\subsubsection{Randomized broadcasting}
\label{ca}

An algorithm which performs broadcasting in multi-hop radio networks without 
collision detection was  developed by Bar-Yehuda, Goldreich and 
Itai~\cite{Bar-YehudaGI92}.
It operates in time $\cO((D+\log n/\epsilon)\log n)$, and succeeds with 
probability $1-\epsilon$. 
The number  $\epsilon>0$ is a parameter, which is part of the code. 

A brief exposition of the algorithm is as follows. 
There are two procedures {\sf Decay} and {\sf Broadcast}.
An essential property of procedure {\sf Decay} is that if node~$x$ has some 
$d$~neighbors, and they all know the message and simultaneously perform
{\sf Decay}$(k)$, where $k\ge 2\lg d$, then node $x$ will hear the message 
by time step~$k$ with the probability greater than~$1/2$.

\begin{myquote}
{\tt procedure}  {\sf Decay}$(k)$; \\
\hspace*{3em}
{\tt repeat}  at most $k$ times \ :\\
\hspace*{6em}
transmit the message ;\\
\hspace*{6em}
toss a coin ;\\
\hspace*{3em}
{\tt until} \ tails come up\ .
\end{myquote}

The broadcasting algorithm itself begins with the source node transmitting 
the message to the nodes reachable from it directly, then all the nodes 
perform procedure {\sf Broadcast}, which consists of procedure 
${\sf Decay}(k)$ iterated about $\log (n/\epsilon)$ times, where 
number~$k$ is about $\log \Delta$. 
A precise description of {\sf Broadcast} is as follows:

\begin{myquote}
{\tt procedure} {\sf Broadcast ;} \\
\hspace*{3em}
set $k = 2\lceil \lg \Delta\rceil$ ; \\
\hspace*{3em}
wait until receiving the message ; \\
\hspace*{3em}
{\tt repeat} $2\lceil \lg (n/\epsilon)\rceil$  times : \\
\hspace*{6em} wait until the time step is a multiple of $k$ ; \\
\hspace*{6em} call {\sf Decay}$(k)$ ; \\
\hspace*{3em}
{\tt end repeat} .
\end{myquote}

The expected number of transmissions performed by the algorithm  is 
$\cO(n\cdot\log(n/\epsilon))$. 
Observe that the information reaching a node might have traversed the network
via paths of different lengths, so the broadcasting algorithm cannot be used
directly to find distances to the source node.

A Monte Carlo algorithm to construct the Breadth-First Search 
(BFS) tree was developed in~\cite{Bar-YehudaGI92}, 
which also finds the distances of each node  to the root.
The algorithm is again based on procedure {\sf Decay}, which now is 
slowed down to increase the probability to detect distances correctly
from the times of arrivals. 

The algorithm refers to local variable {\tt distance}, to store the distance
to the source node, and the current time step given by {\tt time}. 

Each node runs the following procedure BFS:

\begin{myquote}
{\tt procedure} {\sf BFS :}\\
\hspace*{3em}
set $k = 2\lceil \lg \Delta\rceil$ ; \\
\hspace*{3em}
wait until receiving the message ; \\
\hspace*{3em}
set ${\tt distance}
=\lfloor {\tt time}/(k\lceil \lg (n/\epsilon)\rceil)\rfloor$ ;\\
\hspace*{3em}
{\tt repeat} $\lceil \lg (n/\epsilon)\rceil$  times : \\
\hspace*{6em} wait until the time step is a multiple of $k\lceil \lg
(n/\epsilon)\rceil$~; \\
\hspace*{6em} call {\sf Decay}$(k)$ ;  \\
\hspace*{3em}
{\tt end repeat} .
\end{myquote}

The algorithm operates in time $\cO(D\log \Delta\log(n/\epsilon))$ 
with the probability $1-\epsilon$, and is correct with the same probability.

\subsubsection{Deterministic broadcasting}

Deterministic broadcasting algorithms in multi-hop radio networks that have
been presented in the literature are usually {\em oblivious}, in the sense 
that a node performs transmissions at steps known in advance, once 
it receives the message.
A precise definition of such algorithms is as follows.
Let a set of nodes be called a {\em transmission}.
A sequence of transmissions $\langle T_1, T_2,\ldots\rangle$ is called a 
{\em schedule}.
Given a schedule, a broadcasting algorithm can be obtained as follows:
a node $v$ broadcasts the message in step~$i$ if $v$ has received the message
before step $i$ and $v$ is in $T_i$.
This is a natural class of algorithms, which do not make the nodes send
anything prior to receiving the source message. 
All deterministic broadcasting algorithms that we discuss are of this form, 
and the lower bounds we mention are for the algorithms in this class.

Schedules are usually obtained by taking families of sets of nodes 
${\cal F}_1,{\cal F}_2,\ldots,{\cal F}_k$ and making the transmissions
to be elements of these families.
One way to arrange this is by {\em interleaving}, in which we cycle through 
the families, and consecutive selections from a family are also done in 
some cyclic order.
More precisely, in the $i$-th step a set in ${\cal F}_{\ell}$ is selected,
where $i\equiv \ell \bmod k $.
Similarly, in the $j$-th selection form ${\cal F}_{\ell}$ the element is taken
whose index is congruent to $j$ modulo the size of~${\cal F}_{\ell}$.
Families are needed which have useful combinatorial properties, to guarantee
correctness of broadcasting, and be as small as possible, to provide
efficiency.
This paradigm was first applied by Chlebus,  G\c asieniec,  Gibbons,  
Pelc and  Rytter~\cite{ChlebusGGPR00,ChlebusGGPR02}, where the notion of a selective family 
was introduced.
A family ${\cal F}$ of subsets of $[1..n]$ is {\em $k$-selective\/}
if, for any subset $A\subseteq[1..n]$ such that $|A|\le k$, there is 
$B\in{\cal F}$ such that $|A\cap B|=1$.
The family of singletons is $k$-selective, for any number~$k\le n$,
its size is~$n$.
The algorithm whose transmissions cycle through it is called {\sc RoundRobin},
it performs broadcasting in time $\cO(n^2)$.
Smaller selective families can be defined as follows.
Let $i$ be a positive integer, $i\le \lg n$.
Let $C\subseteq [1..\lceil \lg n\rceil]$ be a set of size~$i$.
Let $D$ be a string of $0$-s and $1$-s of length~$i$.
Define $S(C,D)$ to consist of these nodes whose IDs in binary have 
the bit on the $j$-th position in $C$ equal to the $j$-th element in~$D$, 
for $1\le j\le i$.
Let family ${\cal R}(i)$ consist of all such sets $S(C,D)$.
One can show that ${\cal R}(i)$ is $2^i$-selective.
A deterministic broadcasting algorithm obtained by interleaving 
{\sc RoundRobin} with a certain specific ${\cal R}(j)$, where $j$ depends
on~$n$, was shown in~\cite{ChlebusGGPR00,ChlebusGGPR02} to operate in time $\cO(n^{11/6})$.
The algorithm obtained by interleaving all the families ${\cal R}(i)$
such that $|{\cal R}(i)|=\cO(n)$ was shown in~\cite{ChlebusGOR00} to work in time
$\cO(n^{2-\lambda+\epsilon})$, for any $\epsilon$, where 
$\lambda\cong 0.22709$ is defined by the equation 
$\lambda+H(\lambda)=1$, where~$H$ is the binary entropy function.
It was shown~\cite{ChrobakGR-JA02,ClementiMS03,Peleg2000-manuscript} that for each $k\le n$ there is a 
$k$-selective family of size~$\cO(k\log n)$.
Various selective families, their combinatorial properties and applications to
broadcasting were studied in~\cite{ChlebusGGPR02,ChlebusGOR00,ChrobakGR-JA02,ClementiMS03,DeMarcoP01,Peleg2000-manuscript}.
The fastest currently known deterministic broadcasting algorithm was developed
by Chrobak,  G\c asieniec and Rytter~\cite{ChrobakGR-JA02}, it works
in time~$\cO(n\log^2 n)$ and is based on a selective family, with a certain 
additional property, shown to exist by the probabilistic method.
A constructive deterministic broadcasting algorithm 
operating in time~$\cO(n^{3/2})$ was given by
Chlebus, G\c asieniec, \"{O}stlin and  Robson~\cite{ChlebusGOR00}, it is currently the
fastest constructive algorithm known.

Brusci and  Del Pinto~\cite{BruschiP97}
proved a lower bound $\Omega(D\log n)$, for $D$ up to $\Theta(n)$, 
for deterministic broadcasting in multi-hop radio networks.
Clementi, Monti and Silvestri~\cite{ClementiMS03} showed a similar lower bound
$\Omega(n\log D)$, for $D$ up to $\Theta(n)$.
An impossibility of acknowledgement, while broadcasting 
with no collision detection in general graphs, was showed in~\cite{ChlebusGGPR02}.

\subsubsection{Multiple communication}

Bar-Yehuda,  Israeli and Itai~\cite{Bar-YehudaII93} considered multiple instances 
of point-to-point communication and broadcast.
They developed randomized Las Vegas algorithm to achieve this in the model of
separate messages.
Networks are assumed to be symmetric and nodes know their neighborhood.
Their algorithm begins with preprocessing.
First a leader is elected, by resorting to the algorithm 
described in Subsection~\ref{sim}, then a BFS tree is found, rooted at the
leader. 
The expected time of preprocessing is $\cO((n+D\log n)\log\Delta)$. 
The protocol itself consists of two phases: during the first 
{\sf Collection} phase all the messages are collected at the root, then the
{\sf Distribution} phase follows, during which the messages are sent from the
root towards their destinations.
The idea is to send messages to the root of the BFS tree and then to the 
target nodes along the edges of the tree, pipelining messages 
along the branches of the BFS tree.
A deterministic local acknowledgement is used, which immediately 
confirms a successful receipt of a message along an edge.
A probabilistic analysis is done by interpreting the system as 
a line of queues, each with Bernoulli service and arrival times
(see~\cite{ChlebusDP94-SIAM} for an alternative approach to similar problems).
A set of $k$ point-to-point transmissions require the expected 
time $\cO((k+D)\log \Delta)$.
Similarly, $k$~broadcasts require the expected time 
$\cO((k+D)\log \Delta\log n)$.

The broadcasting algorithm presented in Subsection~\ref{ca} 
is Monte Carlo, but it may be converted to be Las Vegas,
as was shown in~\cite{Bar-YehudaII93}. 
The method is as follows.
Since the size~$n$ is known, choose $\epsilon=1/n$ and build a BFS tree. 
After the tree is expected to have been completed, collect the information 
along its edges from all the nodes in the tree by the procedure 
{\sf Collection} discussed in the preceding paragraph. 
This is done within a certain time interval, and if not successful, 
then the whole algorithm is repeated from the very beginning.

A method to obtain a gossiping algorithm from one performing broadcasting
was developed in~\cite{ChrobakGR-JA02}.
If the broadcasting algorithm works in time $\cO(B(n))$ then the resulting
gossiping algorithm operates in time $\cO(\sqrt{B(n)}n\log n)$.
Applying this method with the broadcasting algorithm from~\cite{ChrobakGR-JA02}
with $B(n)=n\log^2 n$ yields a gossiping algorithm working in time
$\cO(n^{3/2}\log^2 n)$.
Similarly, the constructive algorithm from~\cite{ChlebusGOR00} with $B(n)=n^{3/2}$
gives a deterministic constructive gossiping algorithm working in
time~$\cO(n^{7/4}\log n)$.
A randomized Las-Vegas gossiping algorithm, working in the expected
time~$\cO(n\log^4 n)$ was given in~\cite{ChrobakGR04}.

\subsection{Simulations}
\label{sim}

Bar-Yehuda, Goldreich and Itai~\cite{Bar-YehudaGI-DC91}
developed a simulation of the multiple access channel with ternary feedback
on the multi-hop radio network without collision detection.
The simulation algorithm uses the procedures {\sf Decay} and {\sf Broadcast} 
presented in Subsection~\ref{ca}.
The slowdown of the simulation is 
$B_\epsilon=\cO((D+\log(n/\epsilon))\log\Delta)$, that is, $B_{\epsilon}$ 
is the time it takes to perform a single broadcast of the channel 
on the packet network, with the probability at least $1-\epsilon$, which is
also the probability of correctness of the protocol.
Additionally, an implementation of conflict-detection mechanism was developed 
for a multi-hop radio network without conflict detection, in time 
$\cO(\log(1/\epsilon)\log\Delta)$ for one step, and with the probability 
$1-\epsilon$.
It was also shown that any broadcast-channel algorithm operating in 
$t$~steps can be emulated in time $\cO(t(\log t\log \Delta+B_\epsilon))$ 
on a packet-network, with the probability $1-\epsilon$.
The main application is leader election in time
$\cO((D+\log(n/\epsilon))\log \Delta\log\log n)$,
obtained by a combination of the simulation and Willard's algorithms,
optimality of this algorithm is an open problem.

Alon, Bar-Noy, Linial and Peleg~\cite{AlonBLP92}
developed step-by-step simulations of message-passing networks on packet 
radio networks.
Two models were considered.
In the {\em general model\/} a node may send in one round arbitrary distinct
messages, and in the {\em uniform model\/} only copies of one message.

Consider first the general model.
The results are as follows.
There are graphs that require slowdown $\Omega(\Delta_m^2)$ for any simulation
algorithm.
Schedules of $\cO(\Delta_{{\rm in}}\Delta_{{\rm out}})$ 
rounds can be found by a polynomial (centralized) algorithm. 
A randomized Las Vegas algorithm was developed with slowdown 
$\cO(\Delta_{{\rm in}}\Delta_{{\rm out}}\log(n/\epsilon))$ 
with probability $1-\epsilon$.

Next consider the uniform model.
It was shown that there are graphs that require the slowdown of 
$\Omega(\Delta_m\log \Delta_m)$, for any algorithm.
There is a polynomial time (centralized) constructible schedule
with slowdown $\cO(\Delta_{{\rm in}}\log n)$.
Randomized Las Vegas algorithm was developed with slowdown 
$\cO(\Delta_{{\rm in}}\log(n/\epsilon))$ with probability $1-\epsilon$.

\subsection{Optimality}

There are two relevant issues concerning optimality of 
randomized communication.
The first one concerns the question whether randomness can improve the
performance of protocols, as compared with purely deterministic ones.
The second task is to find an optimal randomized communication
algorithm in the class of randomized ones.

The problem of broadcasting in ad-hoc radio networks 
is an example of a problem for which randomness can cause an exponential 
speedup in its complexity.
A lower bound $\Omega(n)$ for deterministic algorithms performing 
broadcasting was shown in~\cite{Bar-YehudaGI92}.
It holds even if the networks are restricted to be of diameter equal 
to~$3$ and of especially simple form:
such a network of $n$~nodes consists of three layers, the first layer contains 
only the source, the second layer contains the remaining nodes except one 
{\em sink}, which is the only node in the third layer.
All the nodes from the second layer are reachable from the source,
but only a proper subset of the nodes in the second layer are neighbors
of the sink.
Notice that the algorithm presented in Subsection~\ref{ca} 
performs broadcasting in the expected time $\cO(\log n)$ on such a network,
what gives an exponential gap.

To assess optimality of the broadcasting algorithm (of Section~\ref{ca}),
in the class of randomized algorithms, we may compare its expected performance 
with the following two lower bounds. 
The first one is $\Omega(\log^2 n)$, it holds for a family of graphs 
of diameter two, it was proved by Alon, Bar-Noy, Linial and 
Peleg~\cite{AlonBLP91}. 
The other lower bound is $\Omega(D \log (n/D))$, it was proved by
Kushilevitz and Mansour~\cite{KushilevitzM-SICOMP98}.
It follows that the expected complexity $\cO(\log^2 n+ D\cdot\log n)$ 
of the broadcasting algorithm is optimal for networks of small diameter. 
Namely, the upper bound becomes $\cO(\log^2 n)$ for networks 
of diameter $D=\cO(\log n)$, which is optimal by the first lower bound, 
and if the diameter~$D$ satisfies $D=\cO(n^a)$, for a constant $a>0$, 
then the upper bound becomes $\cO(D\log n)$, which is again
optimal by the second lower bound.

The lower bound $\Omega(\log^2 n)$  is actually in the following strong form: 
there is a family of graphs, with $n$~nodes and of diameter two, for which any 
algorithm requires time $\Omega(\log^2 n)$.
It follows that the bound holds for randomized algorithms and the graph 
may be even assumed to be known.
The proof is by the  probabilistic method.

The lower bound $\Omega(D \log (n/D))$ is of a weaker form: 
for any randomized broadcasting algorithm, and for each numbers $n$ and
suitable~$D$, there exists a network of $n$ nodes and of diameter~$D$, such 
that the expected broadcasting time of the algorithm is $\Omega(D \log (n/D))$.
The bound is valid for a class of algorithms which are arbitrary functions
determining the probability of broadcasting of a node in a step,
depending on both its ID and its history.
The proof is based on the following lemma: for a single-hop radio network
without collision detection, there is a subset of stations such that if exactly
they are active then the expected time till the first successful broadcast is
$\Omega(\log n)$.
The main lower bound is obtained by constructing a layered network, 
in which the nodes are partitioned into layers in such a way 
that two consecutive layers induce a complete binary graph, 
and there are no other edges except for those connecting any two nodes 
in consecutive layers.
The number $D$ of layers is the depth, and the lemma is applied to each layer
of size~$n/D$.

\section{Related work}

In this section we present additional relevant results concerning 
communication algorithms in radio networks. 

G\c asieniec, Pelc and Peleg~\cite{GasieniecPP-JDM01} compared the locally and globally
synchronous models of single-hop radio networks in the context of the
{\em wakeup\/} problem, in which the time when each station joins the protocol
is controlled by an adversary, and the goal of an algorithm
is to perform a successful broadcast as soon as possible.
If the stations have access to a global clock then wakeup can be realized 
in time $\cO(\log n \log(1/\epsilon))$ with the probability at least
$1-\epsilon$, by iterating procedure {\tt Decay}.
If the local clocks are not synchronized, then processors may flip coins 
with the probability~$1/n$ of heads to come up, to decide if to transmit 
in a step, this yields an algorithm operating in time 
$\cO(n\log(1/\epsilon))$ with the probability at least $1-\epsilon$.
Deterministic algorithms in such setting require time $\Omega(n)$,
and deterministic schedules are known to exist achieving 
time~$\cO(n\log^2n)$.

Dynamic aspects of radio networks include not only the mobility of
nodes, but also the possibility of changing the ranges of nodes 
by modifying their power consumption.
To study such aspects, the notion of a radio network needs to be modified,
for instance by assuming that the nodes are in some metric space and 
the reachability relation is determined by ranges that the nodes 
can cover.

A multi-hop network is called {\em power-controlled\/}	if 
hosts are able to change their transmission ranges, and in this way change 
also the topology of the underlying reachability graph.
Adler and Scheideler~\cite{AdlerS98} introduced a general
formal model of power-controlled networks, in which each pair of vertices 
is assigned the lowest transmission power 
that allows to maintain a direct connection between the vertices.
They developed efficient strategies for permutation routing in such networks.
They considered also a special case where the nodes are located in a square
region in the plane, distributed randomly, and the power consumption to
maintain a connection is proportional to the distance.
They showed that permutation routing on such $n$-node networks can be 
performed in~$\cO(\sqrt{n})$ steps, which was then proved to be optimal 
by a matching lower bound $\Omega(\sqrt{n})$.

Diks, Kranakis, Krizanc and Pelc~\cite{DiksKKP02}
developed deterministic algorithms and lower bounds for broadcasting 
in the case of nodes located on a line, the nodes know their coordinates.
Ravishankar and Singh~\cite{RavishankarS95,RavishankarS94} studied broadcasting and gossiping
when stations are  randomly placed on a line.
Kranakis,  Krizanc and  Pelc~\cite{KranakisKP-JA01}
studied deterministic fault-tolerant broadcasting in radio networks with
the stations located either on a line or on a two-dimensional grid, 
some of them faulty.

Kushilevitz and Mansour~\cite{KushilevitzM05} considered communication 
in noisy single-hop networks.
The specific problem they studied is that of computing a threshold function
of the bits stored in the nodes, one bit per node.
The protocols are assumed to be oblivious, so the bits cannot be encoded by
silence/transmission sequences.
Every bit transmitted is flipped randomly and independently over the receiving
stations, due to noise, the probability that a wrong bit is received is
below $1/2$, but this probability is not assumed to be known by the
stations.
Let $n$ be the number of stations.
A protocol is developed which, for a given natural number $k\le n$ and 
$\epsilon>0$, decides if the number of bits equal to~$1$ is at least~$k$ with
the probability at least $1-\epsilon$ by performing $\cO(n)$ transmissions.

The following are some results on centralized off-line 
algorithms computing broadcast schedules, 
the networks are known then, and are inputs to such sequential algorithms.
Chlamtac and Weinstein~\cite{ChlamtacW91} developed a deterministic 
algorithm which finds in polynomial time a schedule giving 
broadcasting time $\cO(D\cdot \log^2)$.
Gaber and Mansour~\cite{GaberM03} showed that for any packet radio network 
there is a schedule of broadcasting giving time~$\cO(D+\log^5 n)$, 
and that it can be found by a deterministic polynomial algorithm. 
Chlamtac and  Kutten~\cite{ChlamtacK85} and Sen and Hun~\cite{SenH97}
showed the NP-completeness of finding optimal broadcasting schedules
in packet radio networks,
in particular it was shown in~\cite{SenH97} that the problem remains NP-complete
even if the nodes are points in the plane and the reachability is 
determined by distances.
Kirousis, Kranakis, Krizanc and Pelc~\cite{KirousisKKP00} studied the problem of
assigning transmission ranges to nodes located on a line 
so as to minimize the total power consumption and preserve strong 
connectedness.
This was later extended by Clementi, Ferreira, Penna, Perennes 
and Silvestri~\cite{ClementiPFPS03}.
Chlamtac and Kutten~\cite{ChlamtacK87} showed the NP-completeness
of minimizing the average time per node during broadcasting when nodes 
are blocked from hearing messages other than those resulting from 
a broadcast algorithm.

\section{Discussion}

Most of the material in this chapter concerns single-hop radio networks.
This is not surprising as they have been investigated since the beginning 
of the 1970's, while the research concerning distributed randomized
communication in ad-hoc data radio networks is much newer and
goes back to the middle of the 1980's.

While the multiple access channel has been investigated over the years, 
first the infinitely-many users model was dominant but later the research 
shifted to the finitely-many users model.
The original popularity of the infinitely-many users model seems to have
stemmed from the belief that it was a clean model to prove a good behavior 
of nice protocols in a situation when the number of stations 
may be arbitrarily large. 
Later on, after it has been shown that such protocols like Aloha and
both the polynomial and exponential backoffs are not stable in this model,
the finitely-many users model became more popular.
Does the existing body of research provide evidence that 
results for the infinitely-many users model are less meaningful in general 
because the model ``distorts'' the true behavior of protocols?
We believe that this is not the case,  the following are some of the reasons.
First, the stability of superlinear polynomial backoff protocols in the 
finitely-many users model simply confirms the stabilizing effect, 
for this class of protocols, of a scenario when messages can 
be queued, as opposed to a more challenging situation when messages 
run protocols on their own.
Secondly, the existence of constant-delay protocols for both of these models 
shows that such ultimate good behavior is possible in both models.

The multiple access channel is widely used in existing distributed systems.
This popularity has created a common belief that certain
specific protocols used, like the one governing conflict resolution in 
Ethernet, and exponential backoff protocols in general,  
are the best to provide simple and efficient communication in local 
area networks.
This is in contrast with recent theoretical work,
let us quote~\cite{HastadLR-SICOMP96}: ``Based on our analysis, it would appear that the 
most popular and well-studied protocols are precisely the wrong protocols.''
This statement is apparently meant to be restricted to backoff protocols.
Only recently protocols with a constant expected message delay have been
found, which are not backoff but still are acknowledgement-based, 
their development is a major progress in our understanding how
to operate the multiple access channel.

To sum this up, issues related to communication in single-hop radio 
networks seem to be quite well understood, due to the intensive research 
effort over the past thirty years. 
Multi-hop radio networks have started to be investigated much later, 
and are a fertile ground for future research.

\subsection{Future work}

There are many technical open problems in radio networks, 
and possible ramifications of known results, which concern both 
deterministic and randomized communication modes.
Some of such problems have been mentioned while discussing specific topics.
We conclude our overview with additional open problems which seem to be 
especially interesting.

Consider first the single-hop networks.
The best known upper bound on capacity of the multiple-access channel 
with ternary feedback is slightly above~$1/2$, and the lower bound of the
best algorithm known is a little below~$1/2$.
It is a natural question to ask how is the optimum capacity related to~$1/2$,
in particular if it is simply equal to~$1/2$.
Next, no stable backoff protocol for the infinitely-many users model is known.
What we know is that the binary exponential backoff is unstable for any arrival
rate and that instability always happens for sufficiently large arrival rate.
It is an open question if there is a backoff protocol that would be stable for
some positive arrival rate in this setting, this question was already posed by
MacPhee~\cite{MacPhee-PhD89}.

Our knowledge about the optimal complexity of broadcasting in
multi-hop radio networks is not complete.
The best lower bound for deterministic algorithms is $\Omega(n \log n)$,
whereas the fastest deterministic algorithm known works in 
time~$\cO(n \log^2 n)$. 
This creates a gap of a multiplicative factor $\log n$.
The upper bound is not constructive, we only know an existence proof by the
probabilistic method.
The best constructive design yields an algorithm of performance
$\cO(n^{3/2})$, so here the gap is much bigger.
Next the class of randomized broadcasting algorithms.
The lower bound $\Omega(D\log(n/D))$ shows that an algorithm of the expected
time performance $\cO((D+\log n/\epsilon)\cdot \log n)$ is optimal provided 
$D=\cO(n^a)$, for a constant $a>0$.
If the diameter~$D$ is close to the size $n$ of the network then the 
question of optimality is open.
And do we know if randomization helps in broadcasting?
For networks of small diameter~$D$, in particular if $D$
is polylog in~$n$, then indeed there is an exponential gap in performance
between deterministic and randomized algorithms.
However if $D$ is $\Theta(n)$, then the upper bound
$\cO((D+\log n/\epsilon)\cdot \log n)$ becomes $\cO(n \log n)$.
As mentioned above, there is a gap between the best upper and lower bounds 
for the deterministic algorithms, and we can see that the best upper 
bound for randomized algorithms in arbitrary networks is within this gap, 
actually it is on its boundary determined by the function $n \log n$.
Hence solving the optimality question for deterministic 
algorithms would also answer the question if randomization helps in 
broadcasting when the diameter is close to the size of a network.

We still know little about multiple communication in the model allowing
combined messages.
The gap between the best known deterministic and randomized algorithms
for gossiping is asymptotically bigger than for broadcasting. 
This may be not necessarily due not to the fact that gossiping is much 
more inherently time demanding, but rather because the known deterministic 
gossiping algorithms are much further from optimal.
What are the optimal time complexities of both deterministic and randomized
distributed gossiping in ad-hoc radio networks are interesting open
questions.

Most of the work done on communication in radio networks assumes the globally
synchronous model.
Investigations of the locally synchronous one include the work on the wakeup
problem~\cite{GasieniecPP-JDM01}, also some of the constant-delay protocols developed
in~\cite{GoldbergMPS-JACM00} were shown to be robust enough to allow stations to stop and
then restart.
Sometimes protocols for the stronger model can be redesigned to be able 
to rely on the local synchrony only without a loss in performance, 
this, for instance, holds for broadcasting, as noticed by Peleg~\cite{Peleg2000-manuscript}.
On the other hand, analysis of stability issues of conflict-resolution
protocols for the multiple-access channel typically relies on the global
synchrony.
Clarifying the extent to which communication protocols need global 
synchronization is an interesting topic of research.

\subsection{Survey information}

General overviews of networking technologies can be found in books by 
Bertsekas and Gallager~\cite{BertsekasG92} and Tanenbaum and Wetherall~\cite{TanenbaumW10}.

Issues of wireless and mobile communication are discussed in 
handbooks edited by Gibson~\cite{Gibson2012-book}, Imielinski and Korth~\cite{ImielinskiK1996-book}, Stojmenovic~\cite{Stojmenovic03-handbook}, and in a book by Pahlavan and Levesque~\cite{PahlavanL05-book}.

For more on the multiple-access channel see books by 
Bertsekas and Gallager~\cite{BertsekasG92}, and by Rom and Sidi~\cite{RomC90-book}.
A special issue on random-access communications~\cite{Massey85-editor}
contains survey articles by Gallager~\cite{Gallager85} and Tsybakov~\cite{Tsybakov85},
which cover the investigations on the multiple access channel up to the 
early 1980's.

Ephremides and Hajek~\cite{EphremidesH98}  gave an overview of 
connections between the information theory and communication in networks,
the paper contains a section on the multiple access channel.
A book by Cover and Thomas~\cite{CoverT06} includes a discussion of 
the multiple access channel from the point of view of the
information theory.

A recent survey by Pelc~\cite{Pelc02-chapter} covers broadcasting in multi-hop
radio networks in detail.

\noindent
\textbf{Acknowledgements:}

\noindent
Thanks are due to Leslie Goldberg and Andrzej Pelc for their comments on a preliminary
version of this work. 

\bibliographystyle{plain}

\bibliography{bogdan,books,energy,networks,other,phd-thesis}

\end{document}